\documentclass[12pt]{article}

\usepackage{epsfig}
\newcommand{\be}{\begin{equation}}
\newcommand{\ee}{\end{equation}}
\newcommand{\bea}{\begin{eqnarray}}
\newcommand{\eea}{\end{eqnarray}}
\newcommand{\bt}{\begin{tabular}}
\newcommand{\et}{\end{tabular}}
\newcommand{\nn}{\nonumber}
\newcommand{\ov}{\overline}
\newcommand{\rt}{\rightarrow}
\newcommand{\lrt}{\leftrightarrow}
\newcommand{\omb}{\omega_b}
\newcommand{\neff}{N_{eff}}

\newcommand{\Nature}{{\it Nature\,}}
\newcommand{\PRep}{{\it Phys. Rep.\,}}
\newcommand{\ApJ}{{\it Astrophys. J.\,}}
\newcommand{\ApJS}{{\it Astrophys. J. Suppl.\,}}

\newcommand{\NP}{{\it Nucl. Phys.\,}}
\newcommand{\PR}{{\it Phys. Rev.\,}}
\newcommand{\PRL}{{\it Phys. Rev. Lett.\,}}

\newcommand{\PL}{{\it Phys. Lett.\,}}

\newcommand{\JHEP}{{\it JHEP\,}}
\newcommand{\etal}{{\it et al.\,}}
\begin{document}
\setlength{\unitlength}{1mm} {\hfill $\begin{array}{r}
                \mbox{DSF 25/2003} \\
                \mbox{astro-ph/0307213}
\end{array}$}\vspace*{1cm}

\begin{center}
{\Large \bf Present status of primordial nucleosynthesis after
WMAP: results from a new BBN code}
\end{center}

\bigskip\bigskip

\begin{center}
{\bf A. Cuoco}, {\bf F. Iocco}, {\bf G. Mangano}, {\bf G. Miele},
{\bf O. Pisanti}, and {\bf P.D. Serpico}
\end{center}

\vspace{.5cm}

\noindent {\it Dipartimento di Scienze Fisiche, Universit\'{a} di
Napoli "Federico II", and INFN, Sezione di Napoli, Complesso
Universitario di Monte Sant'Angelo, Via Cintia, I-80126 Napoli,
Italy}
\bigskip\bigskip\bigskip

\begin{abstract}
We report on the status of primordial nucleosynthesis in light of
recent results on CMB anisotropies from WMAP experiment.
Theoretical estimates for nuclei abundances, along with the
corresponding uncertainties, are evaluated using a new numerical
code, where all nuclear rates usually considered have been updated
using the most recent available data. Moreover, additional
processes, neglected in previous calculations, have been included.
The combined analysis of CMB and primordial nucleosynthesis
prediction for Deuterium gives an effective number of relativistic
degrees of freedom in good agreement with the simplest scenario of
three non degenerate neutrinos. Our findings seem to point out
possible systematics affecting $^4He$ mass fraction measurements,
or the effect of exotic physics, like a slightly degenerate relic
neutrino background.
\end{abstract}
\vspace*{2cm}

\begin{center}
{\it PACS number(s): 98.80.Cq; 98.80.Ft}
\end{center}

\newpage
\baselineskip=.8cm

\section{Introduction}
\setcounter{equation}0
\noindent

The recent results of WMAP collaboration on Cosmic Microwave
Background (CMB) anisotropies \cite{wmap1} perhaps represent both
a change of perspective and a further call for increasing
precision in cosmology. Before their first data release, in fact,
the determination of baryonic matter density $\omega_b = \Omega_b
h^2$ was in the realm of Big Bang Nucleosynthesis (BBN), providing
a result which typically is a compromise between the values
preferred by primordial Deuterium and $^4He$ experimental
determinations (see for example \cite{emmpjhep}). The very
accurate determination which is now available from CMB, $\omega_b
= 0.024 \pm 0.001$ \cite{wmap2}, is an independent piece of
information which may be used as a $prior$ in BBN theoretical
analysis. This has dramatic consequences in determining the
internal consistency of the standard picture of light nuclei
formation and/or the role of systematics in the experimental
measurements. As we will discuss in the following, the Deuterium
(density fraction with respect to $H$) experimental result of
\cite{deuterium}, $X_D=(2.78^{+0.44}_{-0.38})\cdot 10^{-5}$, is in
quite a fair agreement with the theoretical prediction using the
CMB value for $\omega_b$. This is a remarkable success of standard
BBN, since $X_D$ is a rapidly varying function of $\omb$. The
status of the other two observed nuclei, $^4He$ and $^7Li$, is
less satisfactory. The source for the tension between theory and
experiments, though not so serious at the moment, may be due to
systematics in the reconstruction of the primordial value from
observations or rather to some more exotic features of the BBN
theoretical scenario, as extra relativistic degrees of freedom or
neutrino chemical potential.

To clarify all these issues it is therefore necessary at this stage to
make an effort to further improve the accuracy of theoretical
predictions, as well as to have new observation and measurement
campaigns. In the recent few years some steps have been made in this
direction: the accuracy of the neutron to proton ratio at decoupling
\cite{massnp}-\cite{emmpnp2} and the numerical stability of the
original public BBN code \cite{kawano} have been improved, the
uncertainties of the theoretical results due to experimental errors on
the several nuclear rates entering in the reaction network \cite{flsv}
have been quantified. This network, partially updated \cite{cvc} with
the recent NACRE compilation \cite{nacre}, traces back to the original
analysis of \cite{fowler} and, more recently, of \cite{skm,cyburt}. In
the last decade, however, many new experimental and more precise data
have been obtained for relevant nuclear reactions in the energy range
of interest for BBN. It is therefore timely to perform a complete and
critical review of the whole nuclear network, in order to refine the
input values for the rates and, perhaps more crucial, to reduce the
theoretical uncertainties on nuclide yields (mainly D and $^7Li$). In
Section 2 we summarize the main results of such an analysis. All
nuclear rates already present in the code of Wagoner-Kawano
\cite{kawano} have been updated using the most recent available data.
Moreover additional processes, neglected in previous calculations,
have been included in a new version of our BBN code. Relevant examples
of updated reactions and new added ones are discussed in this paper,
though a complete overview is in progress \cite{summabibiennae}.

In the following we will be mainly interested in determining
$\omega_b$ and the effective number of relativistic degrees of
freedom, $\neff$, defined as \be \rho_R = \left[ 1+\frac 78
\left(\frac{T_\nu}{T_\gamma}\right)^{4} \neff \right]
\rho_\gamma\, , \label{neff} \ee with $\rho_\gamma$ the photon
contribution to the total relativistic energy density, $\rho_R$.
In the standard scenario three active massless, non degenerate,
neutrinos are the only additional relativistic degrees of freedom.
This results in $\neff=3.04$ at the CMB epoch, which takes also
into account the small entropy release to neutrinos during the
$e^+ -e^-$ annihilation phase and QED plasma corrections
\cite{dhs}-\cite{mmpp}. This effect also slightly modifies BBN and
results in an effective $\neff=3.01$ for $^4He$ mass fraction
\cite{dodelson,dodelson2,dhs}. As known from many analysis
\cite{hansenetal}-\cite{barger}, if one considers $\neff$ as a
fitting parameter, CMB alone cannot severely constrain it, though
WMAP data resulted in a sensible improvement of previous bound. We
have performed a CMB likelihood analysis, described in Section 3,
using the ${\it CMBFAST}$ public code \cite{cmbfast} and the WMAP
software facilities to calculate the likelihood functions
\cite{lambda}-\cite{h03}. Assuming that $\neff$ is not changing
between the BBN and CMB epochs, we show in Section 4 that a joint
analysis of CMB data and Deuterium abundance nicely fits in the
simplest standard scenario, with $\neff \sim 3$.

As we mentioned, the present determination of $^4He$ mass fraction
would require a slightly smaller value for $\omega_b$ than what is
suggested by WMAP data. A possible way out to this discrepancy is
of course to reduce the neutron to proton ratio at decoupling,
e.g. to lower the relativistic energy density. It is therefore not
a surprise that a $purely$ BBN likelihood analysis of the $\neff$
- $\omega_b$ parameters, described in Section 5, provides a
preferred value for $\neff$ smaller than three. We conservatively
interpret this finding as a possible sign of systematics in the
$^4He$ measurements. In Sections 4 and 5 the estimates of
primordial abundances are discussed. Deuterium results for
different quasars absorption systems (QAS) distribute around the
theoretical expected value, while there is still an evidence at
2-3$\sigma$ level for primordial $^7Li$ depletion. We also report
the theoretical prediction for $^6Li$ and $^3He$ abundances, which
may represent further tests for BBN.

In Section 6 we perform an updated analysis of the degenerate BBN
scenario. Similar studies are reported in
\cite{hannestad,barger2}. It has been shown in \cite{pastor2} that
for large mixing angle oscillation solution to solar neutrino
problem, chemical potentials of the three active neutrino species
are much more severely bound than what was found in
\cite{hansenetal}, since oscillations tend to homogenize the
lepton asymmetries among the active neutrino flavors. In
particular, the conservative bound on the asymmetry parameters,
$\xi_\alpha \equiv \mu_{\nu_\alpha}/T_\nu$, quoted in
\cite{pastor2} is $|\xi_{\mu,\tau}| \sim |\xi_e| \leq 0.07$. Of
course there is still room for a small neutrino--antineutrino
asymmetry in the universe, which may help in reconciling the
slight discrepancy in the $^4He$ abundance. Remarkably, a tiny
neutrino chemical potential has a great impact on the BBN favorite
value for $\neff$. We finally give our conclusions in Section 7.

\section{A new code for BBN}
\setcounter{equation}0  \noindent

In this Section we briefly describe the main aspects of a new BBN
code realized starting from the original public version of Kawano
\cite{kawano}. This has been achieved in several steps, some being
already described in previous papers \cite{emmpnp1,emmpnp2}. The
main improvement is a complete review and update of all nuclear
reaction rates which enter in the light nuclide network, as well
as, more generally, a critical review of many other processes
which were not considered in the standard BBN network. Some of
these resulted to be not negligible and thus have been included
for the first time. An exhaustive description of the analysis of
all rates included in the code is in progress
\cite{summabibiennae}. Here we only stress the key role of some of
the processes we have either updated using new experimental
results or included for the first time.

The main aspects of the new code can be summarized as follows:

1) {\it Numerical features}

As discussed in \cite{emmpnp2}, numerical resolution of
differential equation describing the time evolution of nuclide
abundance is quite critical. This is because the problem is {\it
stiff}, since the right hand sides of corresponding Boltzmann
equations, being the difference of almost equal forward and
reverse process rates, are typically a small difference of two
large numbers. To deal with this problem our code adopts a NAG
routine \cite{nag} implementing a method belonging to the class of
Backward Differentiation Formulae, instead of a traditional
Runge-Kutta solver.

2) {\it $n \lrt p$ weak rates and $^4He$ mass fraction}

The weak rates converting neutrons and protons are the key inputs
deciding for the $n/p$ ratio at freeze-out and eventually for the
$^4He$ mass fraction. The theoretical estimate for these rates was
improved by considering QED radiative and finite temperature
effects as well as finite nucleon mass
\cite{massnp}-\cite{emmpnp2}, which result in a reduced
uncertainty in $^4He$ mass fraction $Y_p$ at the level of $0.1\%$.
Present statistical (and possibly systematics) uncertainty of the
$Y_p$ experimental determination, quoted at the level of $1\%$,
does not presently suggest any further effort in theoretical
analysis of this issue. We have also updated the value of the
neutron lifetime, $\tau_n =(885.7 \pm 0.8) s$ \cite{pdg}.

3) {\it An updated nuclear reaction network}

The nuclear reactions involved in BBN form quite a complicated
network of order hundred different processes. The original
compilation \cite{fowler} (standard network) was reviewed by
Smith, Kawano and Malaney \cite{skm}. The standard chain was
partially reanalyzed in \cite{cvc} (for a more recent analysis see
Ref. \cite{cyburt}) by extensively using the NACRE nuclear rate
collection. Since then new data became available, as for example
the LUNA collaboration results for $D+p \rt \gamma+ {^3He}$ (see
later) \cite{luna}. It is well known that the uncertainties on
several nuclear rates still represent the main limitation to any
theoretical analysis and prediction of BBN. On the other hand, an
improved precision of BBN theoretical estimates is mandatory, both
to check the consistency of the physical grounds of BBN and to
constraint the values of cosmological parameters. This can be
achieved by a new complete study of all reaction rate network. Our
analysis can be summarized as follows:

\begin{itemize}
\item[i)] all two body strong and electromagnetic reactions
entering in the production/destru\-ction of light nuclei up to
$A=7$ have been all reevaluated, using NACRE estimates for rates
and errors. In particular the key reaction $n+p\lrt \gamma+ D$ has
been studied in details, both in its experimental and theoretical
aspects, since it represents the main $D$ production process,
therefore influencing all the heavier nuclide primordial
production. \item[ii)] for all processes not included in NACRE
catalogue we have either used other available experimental data or
simple theoretical models to get a reasonable estimate for the
rates. \item[iii)] new processes are now included in the code, as
for example the $^7Li + {^3H}\lrt n + {^9Be}$ stripping/pickup
reaction \cite{brune}, and the analogous $^7Li + {^3He}\lrt p +
{^9Be}$, $^7Be + {^3H} \lrt p + {^9Be}$, which may be relevant in
opening new channels for the production of heavier nuclei with
$A>7$. The rate for $^3He + {^3H} \lrt \gamma + {^6Li}$ has been
also considered and added to the code. This rate was usually
neglected since it is assumed that $^6Li$ is mainly produced via
$^4He + D\rt \gamma + {^6Li}$, whose rate is however poorly known
(different low energy non resonant rate estimates disagree for
even one or two order of magnitude). \item[iv)] we have analyzed
the possible role of some three body processes, as $p+n+n \lrt
n+D$, $n+p+p\lrt p + D$, $p+n+n \lrt \gamma+ {^3H}$, and $n+p+p
\lrt \gamma + {^3He}$. In agreement with standard
approach, we checked that these processes can be in fact safely
neglected, being suppressed by the high entropy per baryon during
BBN.
\end{itemize}
The benefit of this analysis is of course a higher accuracy in
several rates. As a general statement we can say that
uncertainties for $^4He$ and $D$ primordial yields are now largely
dominated by experimental errors. In principle, if no systematics
were present in experimental determinations, a careful analysis of
the consistency of the standard BBN scenario would be possible,
providing stringent constraints on cosmological parameters $\omb$
and $\neff$. Further data, mainly on $D-D$ strong reactions and $
^4He + {^3He} \lrt \gamma + {^7Be}$, are nevertheless still highly
desirable.

As an example of our study we here report some of the main new
results on rates included in the new BBN code. This is by no means
a complete overview, which we will present elsewhere
\cite{summabibiennae}, but it is enough to illustrate the main
aspects of our analysis.

\begin{itemize}

\item[a)] {\it $n+p \lrt \gamma +D$}

This reaction is the main process for $D$ production. There are no
many experimental results for the corresponding rate in the energy
range relevant for BBN \cite{s80}. In the previous versions of the
BBN numerical code the corresponding rate estimate \cite{skm} was
based on the measurement of the thermal capture cross-section,
$\sigma_{th}=(0.3326 \pm 0.0006)\,barn$ \cite{hale91}, with an
overall uncertainty on the rate of $7\%$.

\begin{figure}[t]
\bt{cc}
\epsfig{file=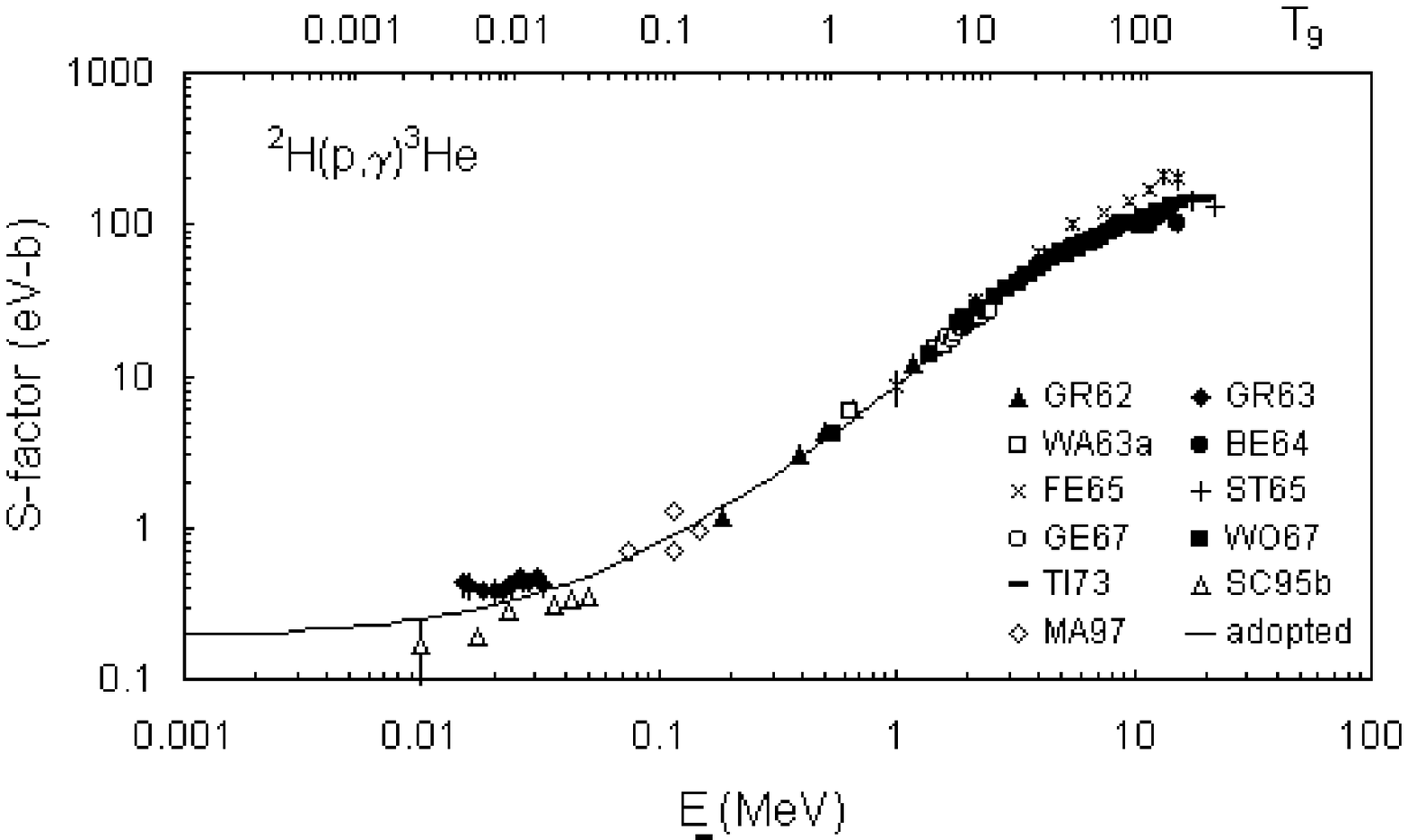,width=7truecm} &
\epsfig{file=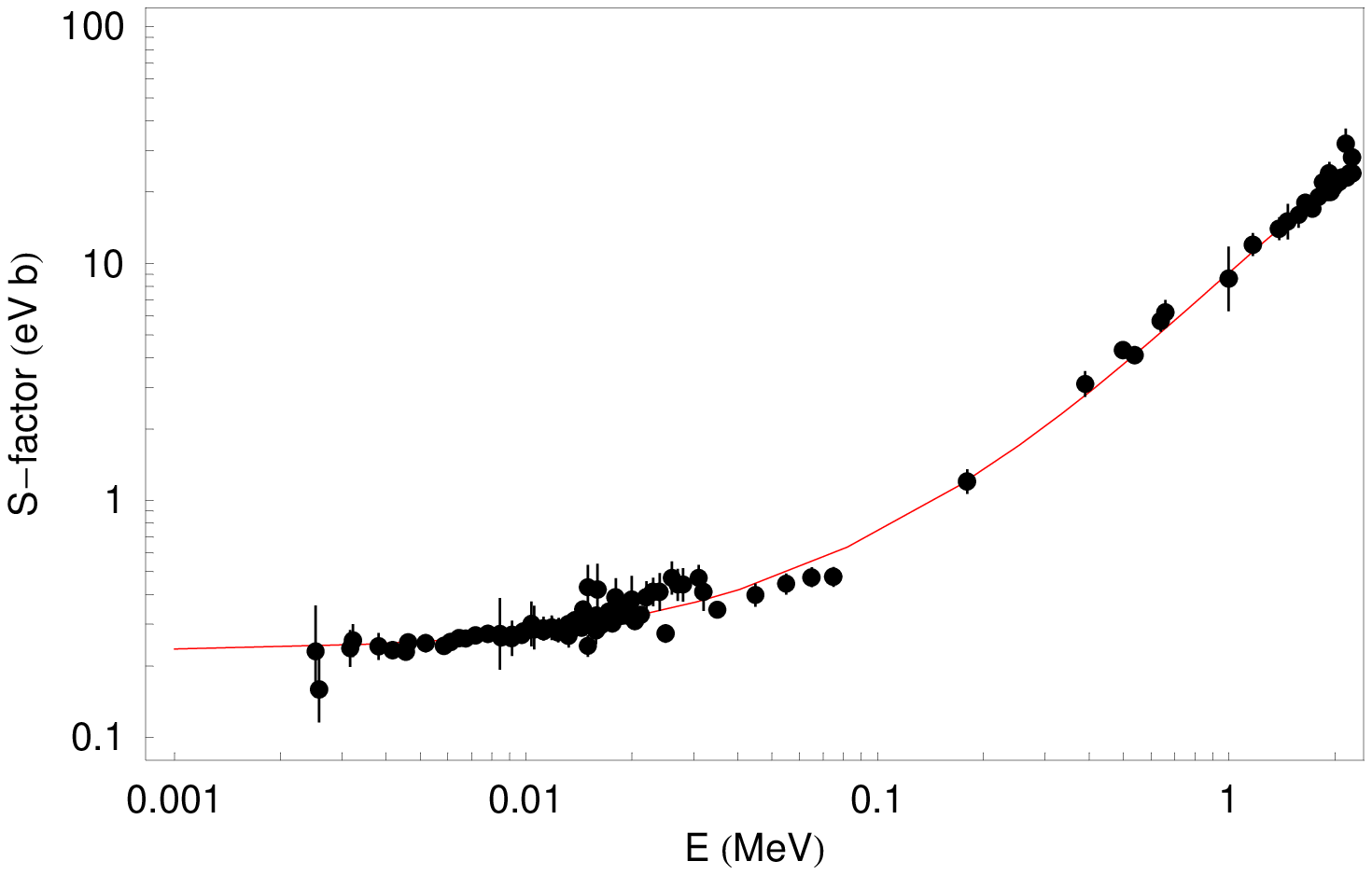,width=7truecm}
\et
\caption{The data and fit for the astrophysical factor of $D + p
\rt \gamma + {^3He}$ from NACRE catalogue (left plot) and with
LUNA data added (the fit of $S(E)$ is with a cubic polynomial).}
\label{fig:nacreluna}
\end{figure}

We have evaluated the rate on the basis of the few low energy data
\cite{pngdata} and mainly using the Effective Field Theory. For
the low energy range, it is possible to describe the process in
terms of an expansion of local operators up to N$^4$LO, with
electromagnetic coupling introduced via gauge principle
\cite{rupak}. The resulting fit of the rate has been then smoothly
linked up with the estimate of \cite{skm}, which is more accurate
for $T \geq 0.1\, MeV$. The overall uncertainty is found to be
$1.2\%$ and $7\%$ for $T\leq 0.1\, MeV$ and $T\geq 0.1\, MeV$,
respectively, which represents a sensible improvement in the
relevant region for BBN.

\item[b)] {\it $ D+p \lrt \gamma +{^3He}$}

This reaction is the main direct source of $^3He$ together with
the strong process $D +D \lrt n + {^3He}$. The corresponding rate
evaluated in \cite{nacre} is based on experimental data which are
accurate in the energy range $0.1 \div 10\, MeV$, while for lower
energies there are two different estimates \cite{s67,s68}. A fit
of the corresponding uncertainty shows a strong temperature
dependence, increasing from $5\%$ at $T \sim 10\,MeV$ up to $30\%$
for $T \sim 0.01\,MeV$.

A sensible improvement of accuracy is provided by a reduced
incompatibility of the two set of measurements \cite{s67,s68}, due
to the presence of systematics in \cite{s67} (see \cite{s69}).
More recently, the LUNA collaboration has performed a new detailed
measurement of this process rate at very low energies \cite{luna},
close to the solar Gamow peak, $E \sim 6.5\cdot 10^{-3}\, MeV$. We
show in Fig.~\ref{fig:nacreluna} the astrophysical factor as
presented in the NACRE catalogue and the fit we have obtained
including the new LUNA data. The rate uncertainty is now reduced
to at most $3.6\%$ (statistical error).

\item[c)] {\it $D+D \lrt n+ {^3He}$} and  {\it $D+D \lrt p+ {^3H}$}

These two reactions are the dominant contribution to,
respectively, $^3He$ and $^3H$ production for $T \leq 0.1\, MeV$.
Moreover they represent a main source of uncertainty in the
theoretical determination of the deuterium final abundance (see
i.e. \cite{flsv}). We have reanalyzed the data collected in
\cite{nacre}, by fitting the (non resonant) astrophysical factor
$S(E)$ with a cubic polynomial (in good agreement with the NACRE
estimate), but the adopted rate has been fitted with a better
accuracy. In our analysis we adopt the NACRE uncertainty. However,
an accurate evaluation of the error budget of the data sets,
presently in progress \cite{summabibiennae}, seems to indicate
that a purely statistical error (below $2\%$) can be attained. In
any case, even such a large reduction of the rate uncertainties
has no relevant consequences on the likelihood analysis, due to
the dominance of the experimental errors.

\item[d)] {\it $^3He + {^3H} \lrt \gamma + {^6Li}$}

This process, modelled with an electric dipole transition
\cite{s70}, is usually neglected in the BBN network, since present
estimates suggest that it contributes only for at most $1\%$ to
$^6Li$ production. This nuclide is in fact mainly synthesized via
the electric quadrupole transition $^4He + D \lrt \gamma +
{^6Li}$. Nevertheless, we have decided to include the $^3He +
{^3H} \lrt \gamma + {^6Li}$ rate for two reasons. First of all the
rate for the process $^4He + D \lrt \gamma + {^6Li}$ is only known
for $E>1\, MeV$ and around the resonance at $E=0.711\, MeV$, while
at lower energies there are only weak upper limits. On the other
hand, the theoretical estimates \cite{nacre} still differ for
order of magnitudes; thus its relative weight in $^6Li$ production
may be smaller than expected. Moreover the $^3He + {^3H} \lrt
\gamma + {^6Li}$ rate itself is possibly affected by a
normalization uncertainty, of the order of a factor 3 (at most)
(see Ref. 8 in \cite{nacre}). Our data fit of the $S$ factor for
$^3He + {^3H} \lrt \gamma + {^6Li}$ with a cubic polynomial shows
a good agreement with the result of \cite{fk90}.

\end{itemize}
We define, as usual, $^4He$ mass fraction, $Y_p$, and relative
density with respect to hydrogen, $X_i~ (i=D,^3He, ^6Li, ^7Li)$,
as
\bea
X_{4He} \equiv Y_p &=& 4 \frac{n_{4He}}{n_b}\, , \nn \\
X_i &=& \frac{n_i}{n_H}\, , \eea with $n_b$ the baryon density. We
do not use $^7Li$ in our analysis to constrain the values of
cosmological parameters (see Section 5 for a discussion of this
nuclide abundance). To compare the experimental values with
theoretical estimates we construct, as customary in literature, a
likelihood function ${\cal L}$ as follows. If $\Omega$ represents
the set of cosmological parameters entering the theoretical model
(basically $\omb$ and $\neff$ and, in the degenerate BBN scenario,
the asymmetry parameter for electron neutrinos, $\xi_e$), and $R$
the set of all nuclear rates $R_k$ involved in the network, the
$joint$ $\chi^2(\Omega)$ function (i.e. $-2~ log{\cal L}$) is
defined by \be \chi^2(\Omega) =
\sum_{ij}\left(X_i^{th}(\Omega,R)-X_i^{exp} \right) F_{ij}
\left(X_j^{th}(\Omega,R)-X_j^{exp} \right)\, , \label{likely} \ee
where $F$ is the inverse of the error matrix and the sum is over
the nuclei considered in the analysis. With $X_i^{th}(\Omega,R)$
and $X_i^{exp}$ we denote the theoretical and experimental
estimate for the {\it i}-nuclide abundance, respectively. In the
following we will be interested in studying either the $D$ or the
joint $D+^4He$ likelihood function, \bea -2\log({\cal L}_D) &=&
\left(X_D^{th}(\Omega,R)-X_D^{exp}
\right)^2 F_{22}\, , \label{likelid} \\
-2\log({\cal L}_{BBN}) &=&
\sum_{i,j=D,4He}\left(X_i^{th}(\Omega,R)-X_i^{exp} \right)F_{ij}
\left(X_j^{th}(\Omega,R)-X_j^{exp} \right)\, , \label{likelibbn}
\eea
For ${\cal L}_D$
\be
F_{22}^{-1} = (\sigma_{22}^{th})^2+ (\sigma_{22}^{exp})^2\, ,
\label{errord}
\ee
while for the joint likelihood $D+\,^4He$ the error matrix is
given by
\be
F^{-1}_{ij}=(\sigma_{ij}^{th})^2+(\sigma^{exp}_{ij})^2\, ,
\label{errormatrix}
\ee
i.e. the sum in quadrature of the (diagonal) experimental error
matrix $(\sigma_{ij}^{exp})^2=\delta_{ij}(\sigma^{exp}_i)^2$ and
the theoretical one $(\sigma_{ij}^{th})^2(\Omega,R)$. The latter,
accounting for the effect on the nuclide abundances of the
uncertainties of the several nuclear rates $R_k$, can be estimated
using the linear propagation method introduced in \cite{flsv}. In
a completely general approach, we consider the quantities
\bea
(\sigma_{ij}^{th})^2(\Omega,R) &\equiv& \frac 14 \sum_k
\left[X_i(\Omega,R_k + \delta R_k^+)-X_i(\Omega,R_k - \delta
R_k^-)\right] \nn \\
& {\times}& \left[X_j(\Omega,R_k + \delta R_k^+)-X_j(\Omega,R_k -
\delta R_k^-)\right]\, , \label{sigmath}
\eea
with $\delta R_k^\pm$ the (temperature dependent) upper and lower
uncertainties on $R_k$, respectively. This differs from the
approach in \cite{flsv} which assumes the existence (in principle
not necessary) of the linear functionals $\lambda_{ik}=\partial
\log X_i(\Omega) /\partial \log R_k$. This only holds for
symmetric and temperature independent relative uncertainties
$\delta R_k/R_k$.

\section{WMAP results and $\omega_b$}
\setcounter{equation}0 \noindent

Since we are interested in using the CMB data in a joint analysis
with the BBN theoretical/experimental ones, the first step in this
study has been to produce our own database and likelihood analysis
of WMAP measurements. We have used the ${\it CMBFAST}$ code
\cite{cmbfast} and the public software provided by the WMAP
collaboration to calculate the likelihood function
\cite{lambda}-\cite{h03}. The temperature and polarization
anisotropies have been evaluated for a set of $\sim 2 \cdot 10^6$
models, varying the following parameters in the reported ranges
with the corresponding steps:
\bea
& \mbox{baryon density} & \,\,\,\, \omb = \Omega_b h^2,\,\,\,\,
0.020 \div 0.029,\,\,\,\, \Delta \omb = 0.0015 \nn \\
&\mbox{cold dark matter density} &\,\,\,\, \omega_{c} = \Omega_{c}
h^2,\,\,\,\, 0.027 \div 0.236,\,\,\,\, \Delta \omega_{c} = 0.011
\nn \\
& \mbox{Hubble parameter}&\,\,\,\, h,\,\,\,\, 0.6 \div 0.8,\,\,\,
\Delta h=0.017 \nn \\
& \mbox{optical depth}&\,\,\,\, \tau,\,\,\,\, 0 \div 0.4,\,\,\,\,
\Delta \tau = 0.05 \nn \\
&\mbox{scalar tilt} &\,\,\,\, n_s,\,\,\,\, 0.8 \div
1.2,\,\,\,\, \Delta n_s = 0.036 \nn \\
&\mbox{relativistic particle density} &\,\,\,\, \neff,\,\,\,\, 0
\div 9,\,\,\,\, \Delta \neff = 1\, .
\eea

\begin{figure}[t]
\begin{center}
\epsfig{file=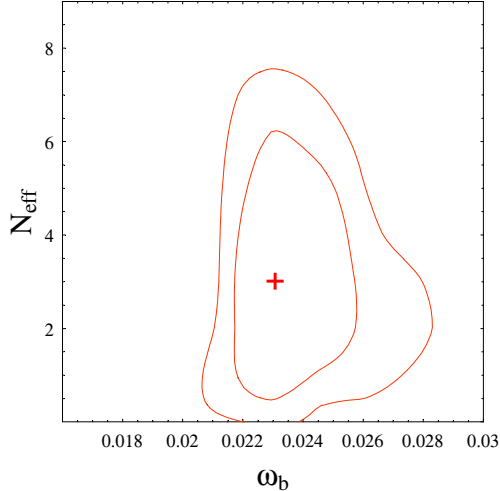,width=7truecm}
\end{center}
\caption{The 68 and 95\% C.L. likelihood contours from WMAP data
in the $\neff-\omb$ plane.}
\label{fig:cmbcontour}
\end{figure}

We only consider spatially flat models, as suggested by the
position of the first acoustic peak, now determined by WMAP with a
rather high accuracy, $l_{p}=220.1 \pm 0.8$ \cite{p03}. We have
also let as a free parameter the overall normalization, which is
eventually fixed by the likelihood analysis. In Table
\ref{cmbtable} we report the best fit values and $95\%$ C.L.
ranges obtained after marginalization over remaining parameters.
Notice that $h$ does not appear in this Table since we consider
its range as a prior.

By using as fitting parameters $\omega_b$, $\omega_c$, $h$, $\tau$
and $n_s$ and fixing $\neff=3.04$ one gets at the best fit
$\chi^2=1431$ for $1342$ d.o.f. . The introduction of $\neff$ as a
new free parameter lowers the value of $\chi^2$ to $1430$ for
$1341$ d.o.f., not significantly improving the goodness of fit.
This is in perfect agreement with the results of \cite{pastor}.

\begin{table}[b]
\begin{center}
\bt{|c|c|c|c|c|c|} \hline \hline & $n_s$ & $\omega_c$ & $\omega_b$
& $\tau$ & $N_{eff}$ \\
\hline\hline
marginalized $(2 \sigma)$
 & $0.98^{+0.11}_{-0.09}$ & $0.11^{+0.09}_{-0.06}$
 & $0.023^{+0.004}_{-0.002}$ & $0.12^{+0.22}_{-0.09}$ & $2.6^{+3.9}_{-2.4}$ \\
\hline
best fit
 & $0.982$ & $0.115$ & $0.023$ & $0.15$ & $3.0$ \\
\hline
\et
\end{center}
\bigskip
\caption{The best fit values and the $(2 \sigma)$ bounds for
marginalized distributions are here reported for the fit analysis
of WMAP data. } \label{cmbtable}
\end{table}

\begin{figure}[t]
\bt{cc}
\epsfig{file=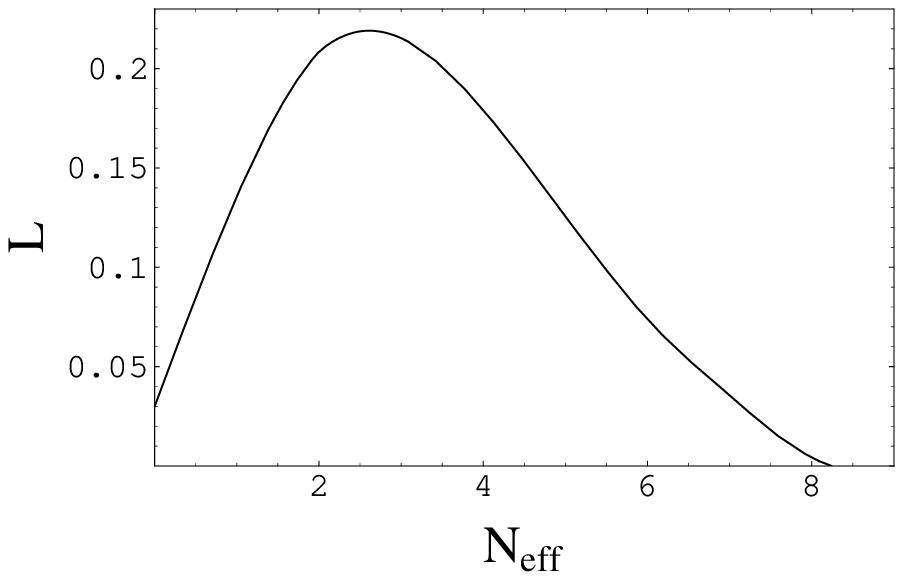,width=7truecm} &
\epsfig{file=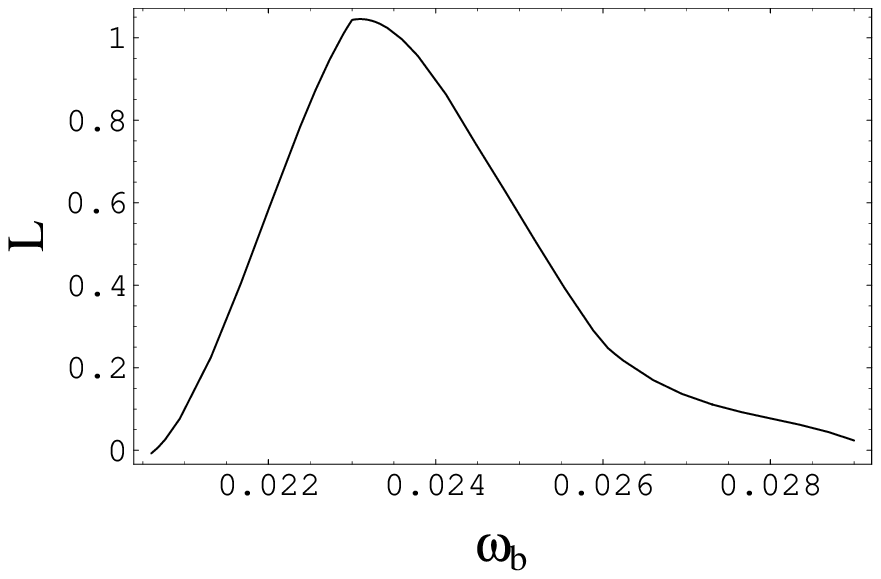,width=7truecm}
\et
\caption{The marginalized likelihoods versus $\neff$ and $\omb$.}
\label{fig:cmbmarg}
\end{figure}

The results are shown in
Fig.~\ref{fig:cmbcontour}-\ref{fig:cmbmarg}. After marginalization
we get $\omb = 0.023^{+0.002}_{-0.001}$ and $\neff =
2.6^{+2.0}_{-1.5}$, at $68\%$ C.L., in good agreement with similar
results obtained in the recent literature
\cite{pastor}-\cite{hannestad}. Compare, for example, our Fig. 3
for $\neff$ with the red curve in Fig. 1 of Ref. \cite{pastor}. We
note that the range obtained in \cite{barger} is instead slightly
larger.

\begin{figure}[p]
\begin{center}
\bt{cc}
\epsfig{file=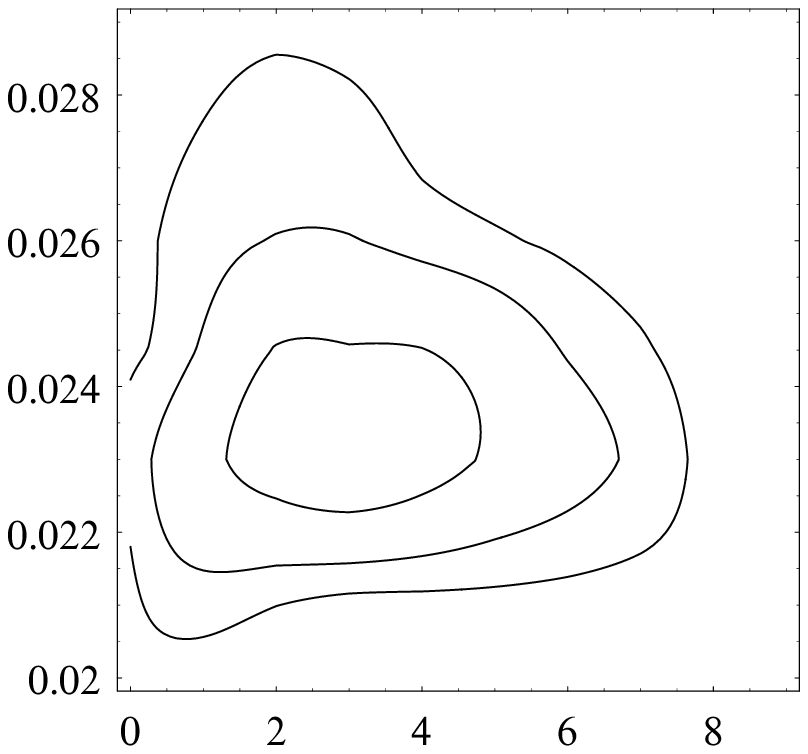,width=4truecm} & \\
\epsfig{file=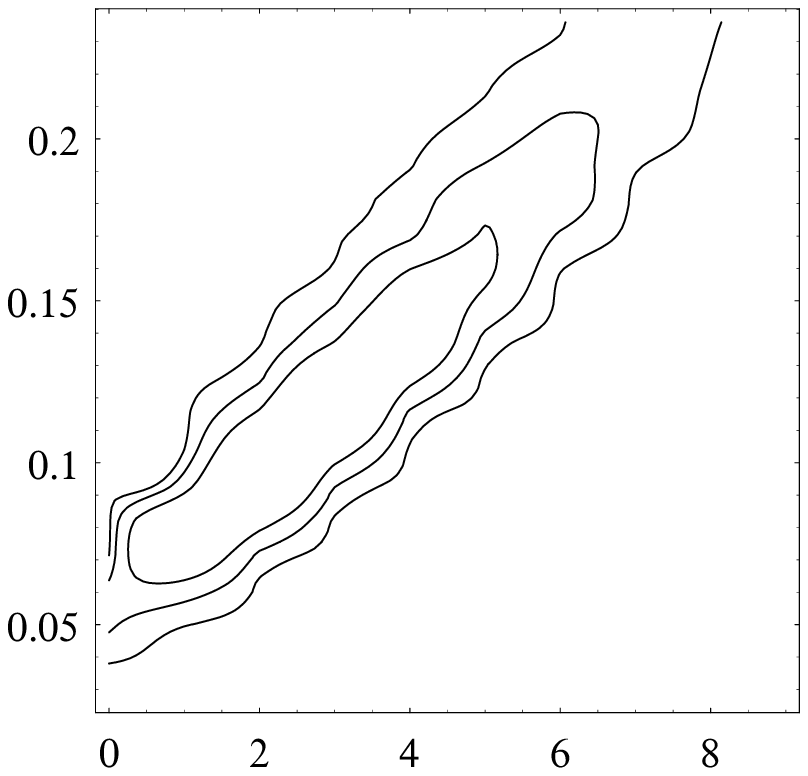,width=4truecm} & \epsfig{file=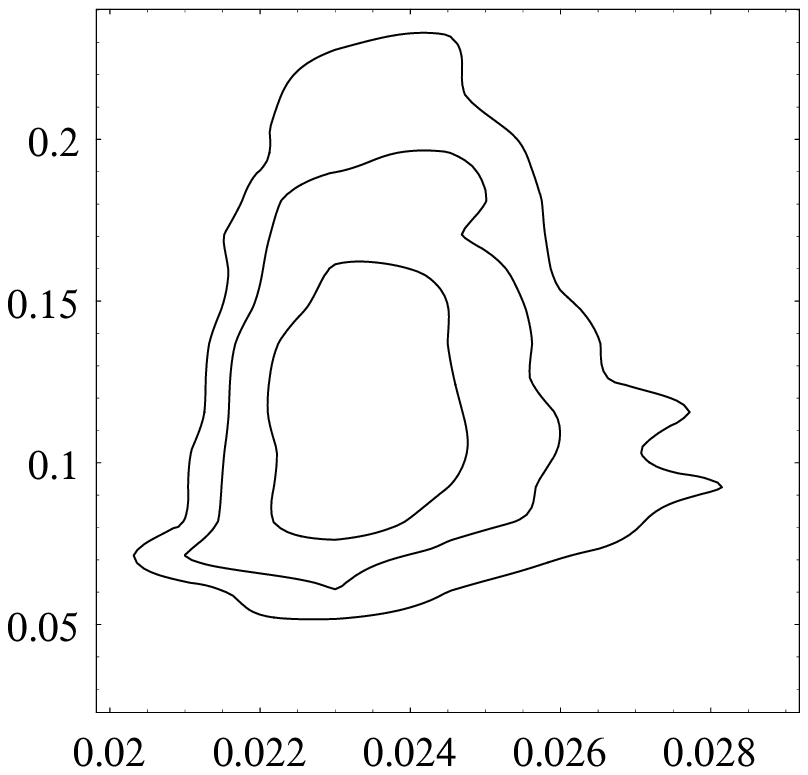,width=4truecm} \\
\epsfig{file=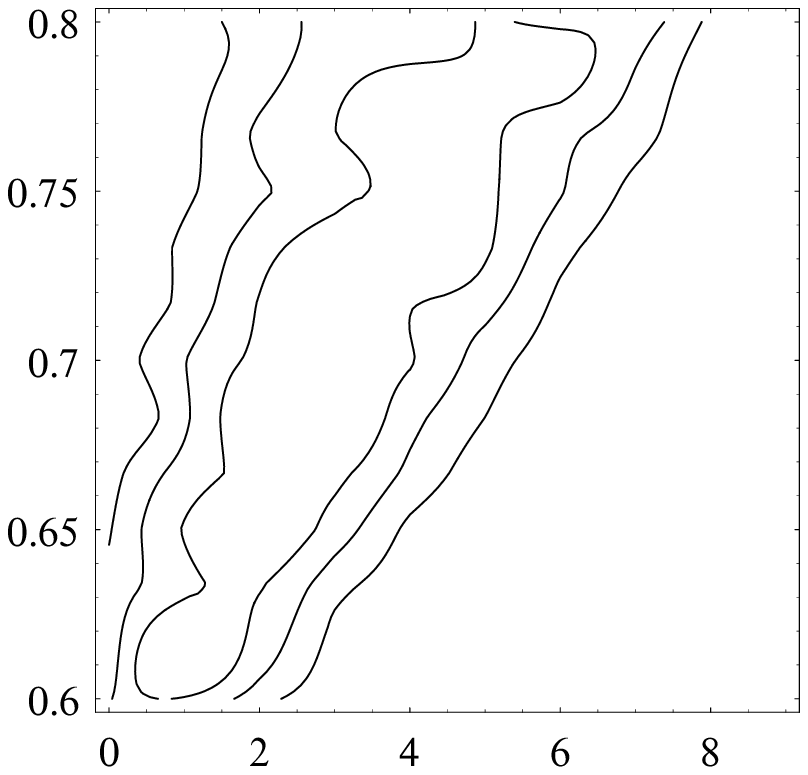,width=4truecm} & \epsfig{file=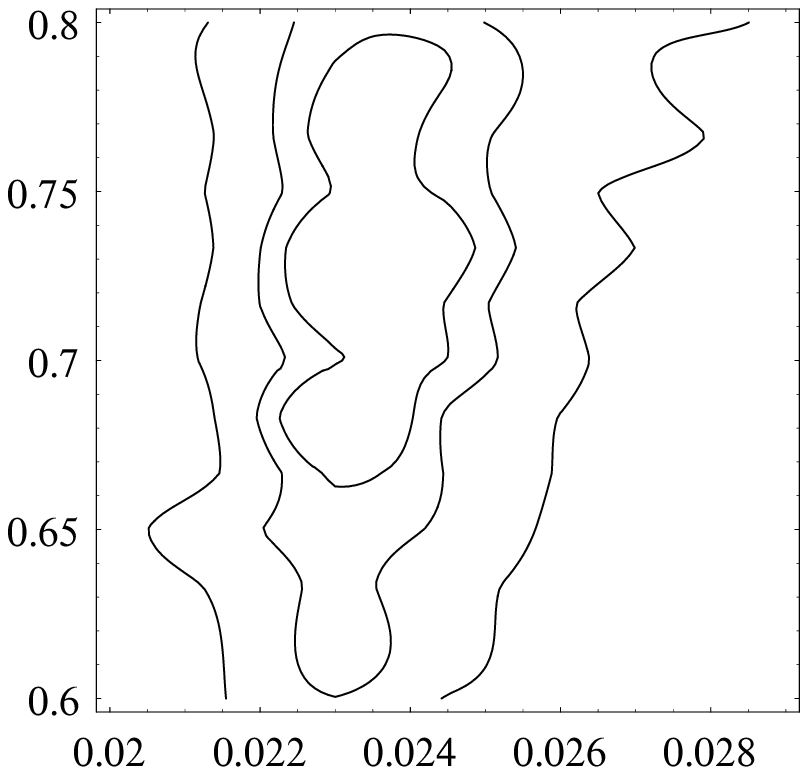,width=4truecm} \\
\epsfig{file=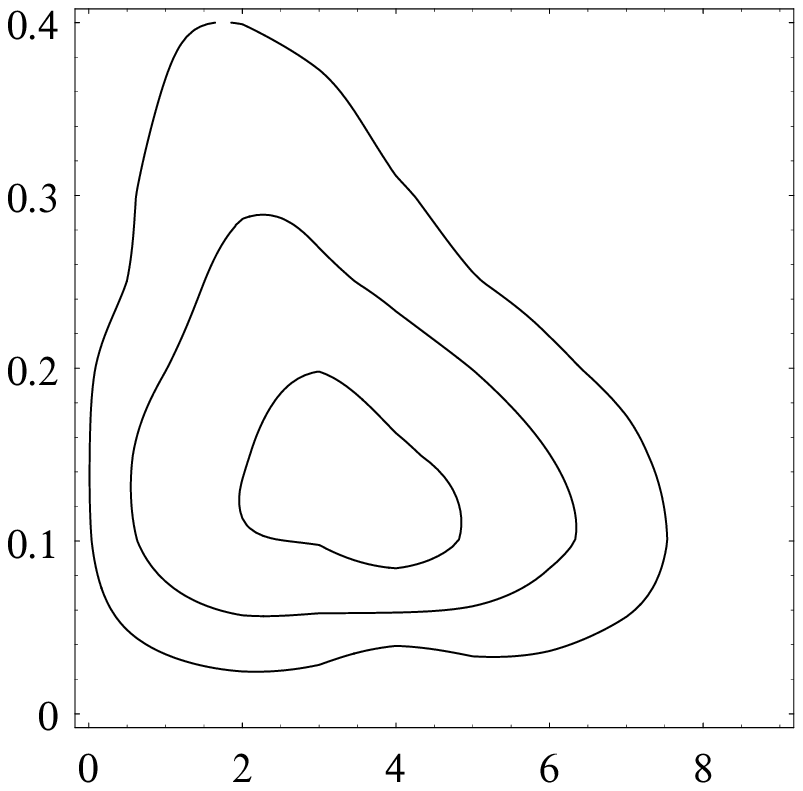,width=4truecm} & \epsfig{file=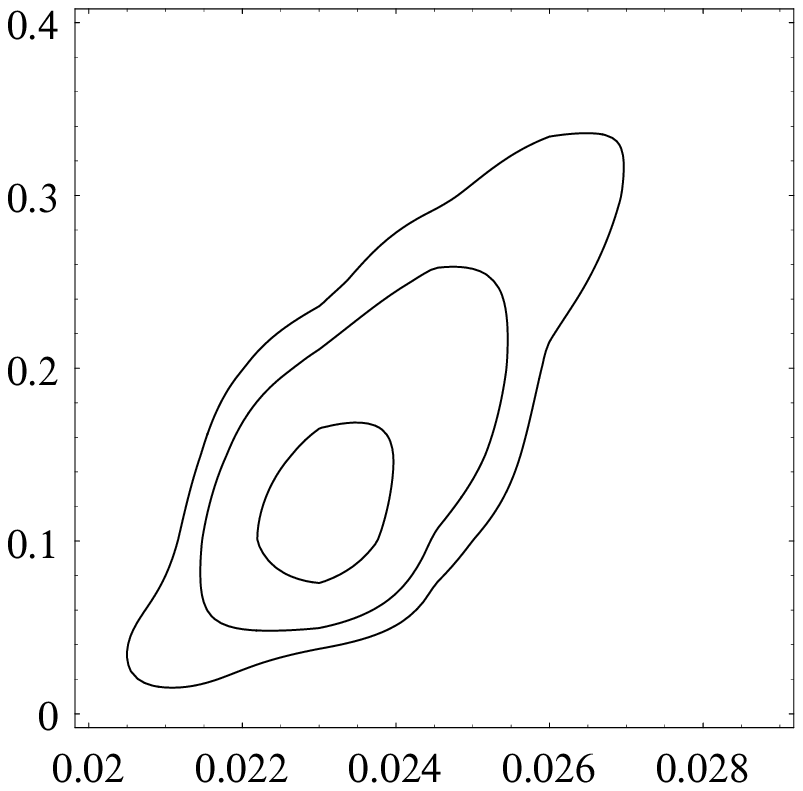,width=4truecm} \\
\epsfig{file=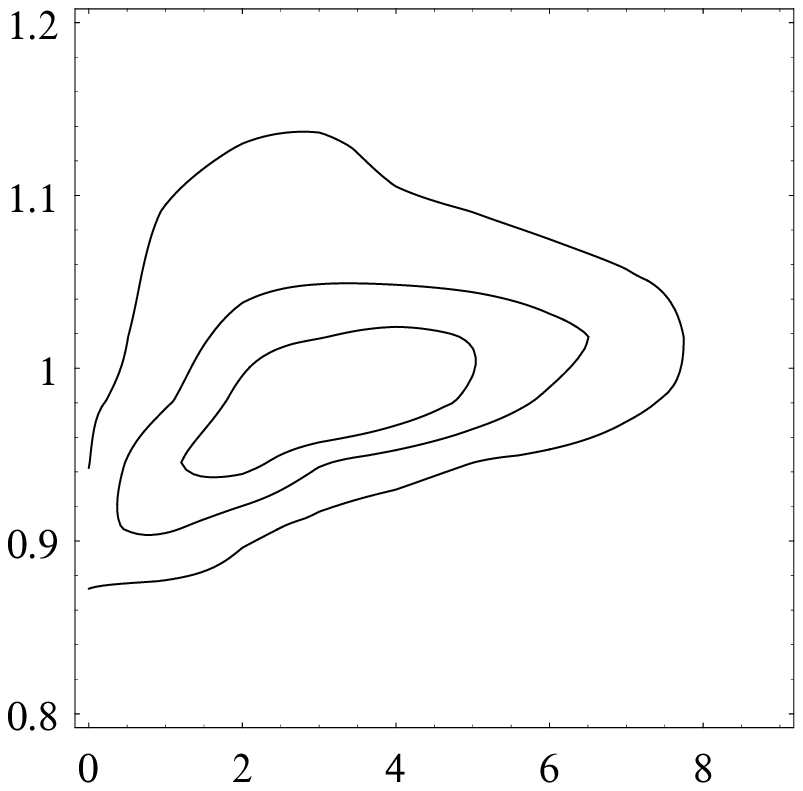,width=4truecm} &
\epsfig{file=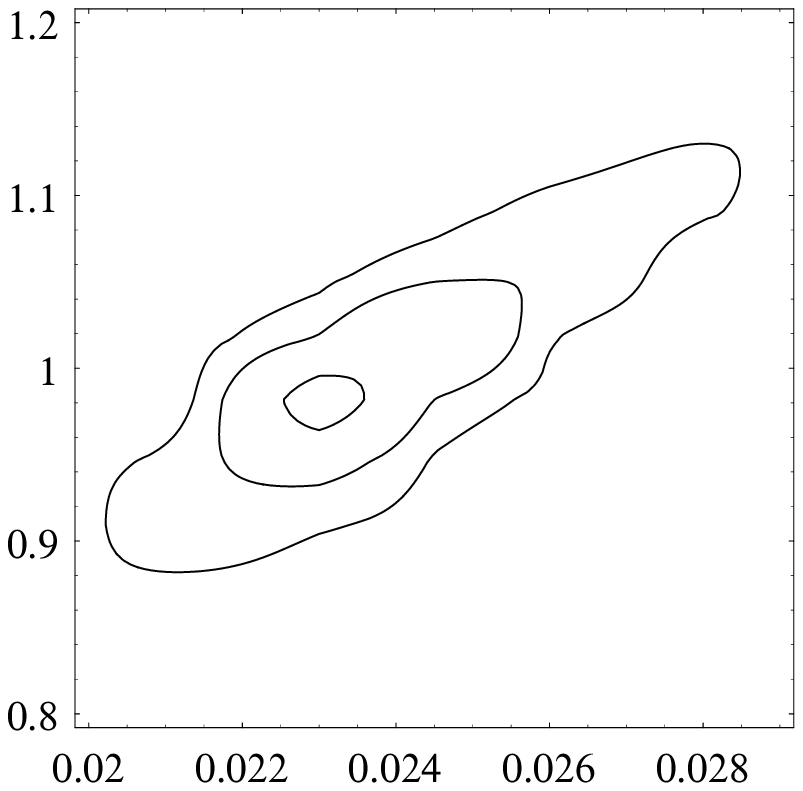,width=4truecm}
\et
\end{center}
\caption{Arbitrary levels of the bidimensional marginalized
likelihood contours: $x$-axis in first (second) column corresponds
to $\neff$ ($\omega_b$), from top to bottom $y$-axis corresponds
to $\omega_b$, $\omega_c$, $h$, $\tau$ and $n_s$, respectively.}
\label{fig:cmb2}
\end{figure}

For completeness we report in Fig.~\ref{fig:cmb2} bidimensional
likelihood contours in all planes involving at least one of the
parameters $\omega_b$ , $\neff$. From this figure we see how CMB
data alone cannot break the degeneracies $\neff$-$h$ and
$\neff$-$\omega_c$, which are the main sources of uncertainty in
the determination of $\neff$. Removing these degeneracies requires
other data sets than CMB. Notice that the ripples in the contours
are an artefact of the finite size in the parameter lattice.

\section{CMB+BBN(Deuterium) analysis}
\setcounter{equation}0
\noindent

Deuterium number fraction, $X_D$, is critically depending on the
baryon content of the universe. It is therefore worth starting our
analysis of the consistency of the CMB + BBN standard scenario by
discussing how WMAP determination of $\omb$ fits with the
experimental determination of $X_D$.

The  estimate of $X_D$ is affected by a theoretical uncertainty
which is dominated by a small number of nuclear rate
uncertainties, mainly the processes which produces either $^3H$ or
$^3He$ out of Deuterium. Varying all the rates $R_k$ which enter
the network with $1\sigma$ errors $\delta R_k^\pm$ (see Section
2), we get from our code, for $\omb=0.023$, $\sigma^{th}_{22}=0.15
\cdot 10^{-5}$. In Table \ref{table:contrsigmad} we also show the
relative contribution of the main processes to this result, in
percentage.

Our previous estimate for the same parameter was $\sigma^{th}_{22}
= 0.21\cdot 10^{-5}$ \cite{emmpnp2}. The main source of this
improvement is due both to the more precise data which are now
available on $D + p \lrt \gamma + {^3He}$ and the reanalysis of
the $D-D$ rates. Considering $\sigma^{th}_{22}$ and the allowed
range for $\omb$ obtained in Section 3, $0.022\leq \omb\leq
0.025$, we get, for the standard value $\neff=3.01$, $X_D=
(2.55^{+0.22}_{-0.34})\cdot 10^{-5}$. This theoretical prediction
nicely fits, within the errors, with the most recent experimental
results of \cite{deuterium}, $X_D= (2.78^{+0.44}_{-0.38})\cdot
10^{-5}$. The latter is the average of several measurements of
$DI/HI$ column ratio in different QSO absorption systems at high
red-shift.

\begin{figure}[t]
\begin{center}
\epsfig{file=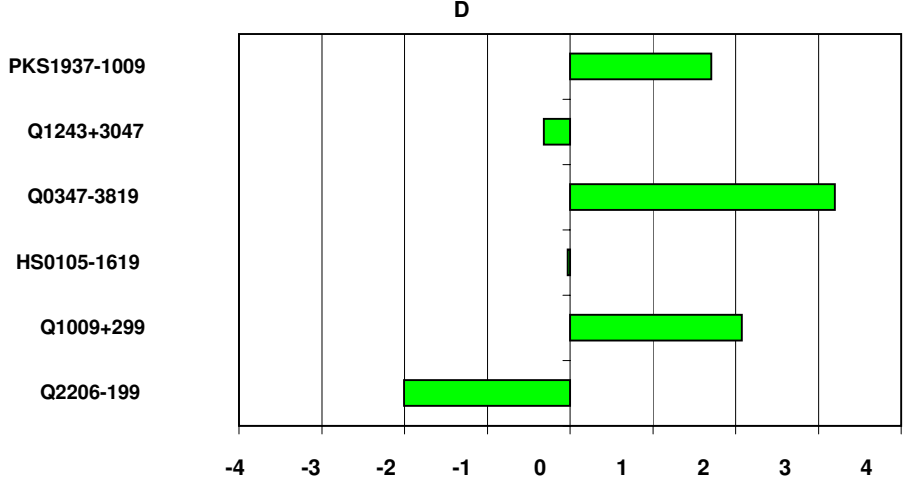,width=12truecm}
\end{center}
\caption{The pulls (see text) of QSO D measurements with respect
to the theoretical prediction for $\neff=3.01$ and $\omb=0.023$,
in units of $\left((\sigma^{th}_{22})^2 + (\sigma^{exp}_2)^2
\right)^{1/2}$.}
\label{fig:pulld}
\end{figure}

\begin{table}[b]
\begin{center}
\bt{|c|c|} \hline\hline rate  $R_k$ & $\delta
(\sigma^{th}_{22k})^2/(\sigma^{th}_{22})^2$(\%) \\
\hline\hline $D+D\lrt n+{^3He}$ & 73 \\ $D+D\lrt p+{^3H}$ & 23 \\
$D+p\lrt \gamma+{^3He}$ & 3
\\ $p+n\lrt \gamma + D$ & 0.6 \\
\hline\hline
\et
\end{center}
\bigskip
\caption{The main contributions to $(\sigma^{th}_{22})^2$, in
percentage, for $\neff=3.01$ and $\omb=0.023$.}
\label{table:contrsigmad}
\end{table}

In Fig.~\ref{fig:pulld} we report the deviations of the
experimental values of $X_D$ with respect to the theoretical
prediction of standard BBN, in unit of the combined error due to
$\sigma^{th}_{22}$ and $\sigma^{exp}_2$ summed in quadrature
(pulls). In spite of the fact that unidentified effects may be
still affecting some of the data, it is nevertheless interesting
to note that, as far as Deuterium is concerned, the standard BBN
scenario with three neutrinos and WMAP result for $\omb$ is quite
consistent.

\begin{figure}[t]
\bt{cc}
\epsfig{file=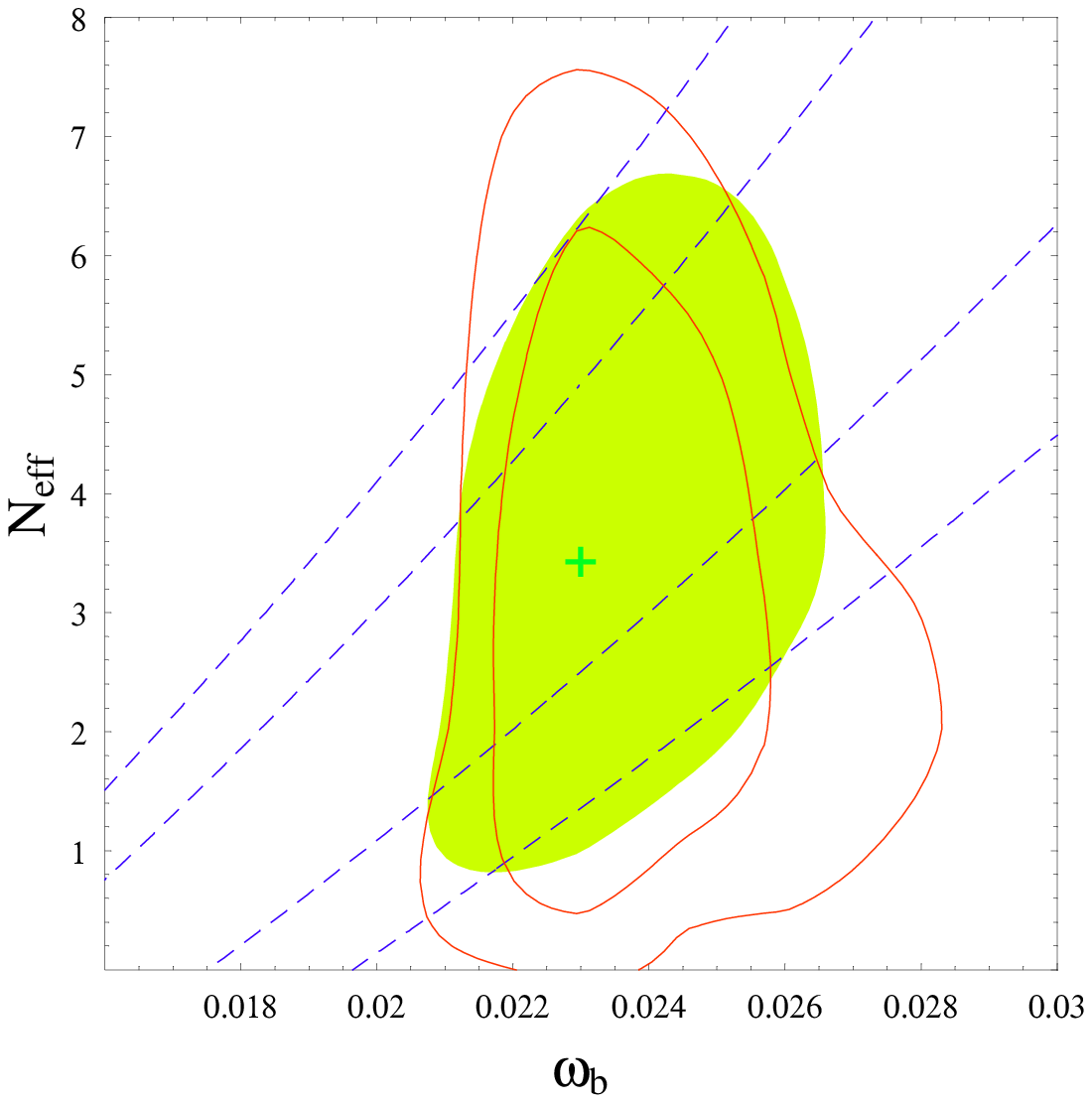,width=7truecm} &
\epsfig{file=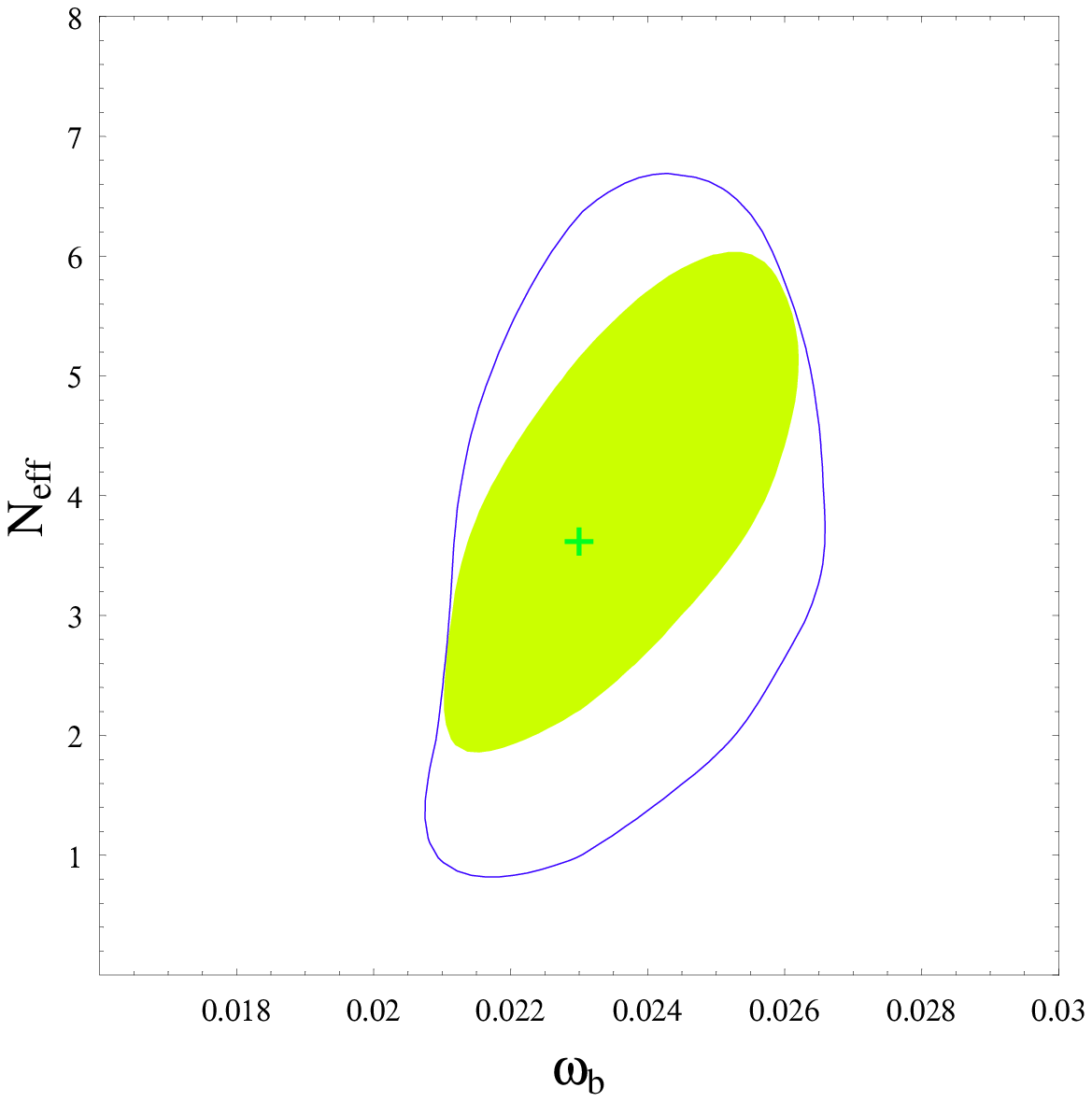,width=7truecm}
\et
\caption{Left plot: the $68$ and $95\%$ C.L. contours in the
$\omb-\neff$ plane for WMAP (solid lines). The dashed lines are
two arbitrary levels of the BBN Deuterium likelihood. The joint
$95\%$ C.L. $CMB+BBN$ region is shown as the filled area. Right
plot: the joint $95\%$ C.L. contour of the left plot (solid line)
is compared to the same contour with a total error on $X_D$
improved by a factor 2 (filled area).}
\label{fig:likelihoodcmbd}
\end{figure}

To analyze this point more quantitatively we have performed a
likelihood study using the WMAP data and BBN Deuterium abundance
only. The likelihood function for $D$ has been constructed as
described in Section 2. The $\neff$ and $\omb$ dependence of
$\sigma^{th}_{22}$ has been fully taken into account. The result
is shown in Fig.~\ref{fig:likelihoodcmbd} (left plot), from which
we see the good agreement of a standard BBN with three active
neutrinos with the WMAP determination of $\omb$.
The product likelihood CMB+BBN is peaked at the values
$\omb=0.023$ and $\neff=3.4$. Likelihood marginalization gives, at
$95\%$ C.L., $\omb = 0.023^{+0.003}_{-0.002}$ and $\neff =
3.6^{+2.5}_{-2.3}$, which give at 1$\sigma$ $X_D=
(2.74^{+0.46}_{-0.51})\cdot 10^{-5}$,
$Y_p=0.256^{+0.014}_{-0.015}$ and $X_{7Li}=
(4.7^{+1.5}_{-1.3})\cdot 10^{-10}$. As expected, $\neff$ is not
severely constrained either by $X_D$ or from CMB: with present
uncertainties on Deuterium, the $95\%$ C.L. interval corresponds
to $1.3 \leq \neff \leq 6.1$. Previous values, when comparisons
are possible, are in nice agreement with the results reported in
\cite{pastor,hannestad}. A more stringent bound on $\neff$ could
be provided by $^4He$ mass fraction. We will discuss this in more
details in the next Section. It is however interesting to see
whether the value of $\neff$ could be more precisely fixed by
Deuterium+WMAP if new and more precise measurements from QSO were
available. We show in Fig.~\ref{fig:likelihoodcmbd} (right plot)
the results of a simulation with an error reduced by a factor two,
$\sqrt{(\sigma^{th}_{22})^2 +(\sigma^{exp}_2)^2} \simeq 0.2\cdot
10^{-5}$. In this case, repeating the same analysis, we obtain
$2.3\leq \neff \leq 5.7$, still a broad range, but not much wider
than what is obtained using a customary BBN likelihood analysis of
both $D$ and $^4He$.

\section{BBN(Deuterium+${^4He}$) analysis}
\setcounter{equation}0 \noindent

The theoretical estimates for $^4He$ mass fraction $Y_p$ and
$^7Li$ number density $X_{7Li}$ show a slight disagreement with
the corresponding experimental results. For $\neff=3.01$ and using
the WMAP result for $\omb$ we get $Y_p=0.2483^{+0.0008}_{-0.0005}$
and $X_{7Li}=(4.9^{+1.4}_{-1.2})\cdot 10^{-10}$. The source of
uncertainties for these two nuclei is as follows. For $Y_p$ the
propagated uncertainty due to nuclear rates, $\sigma^{th}_{44}=
0.0003$, is of the same order of magnitude of the one due to the
variation of $\omb$. Notice how the smaller error on neutron
lifetime, $\Delta\tau_n=0.8\, s$, has strongly reduced the effects
on $(\sigma^{th}_{44})^2$ of weak $n \lrt p$ processes. Concerning
the $X_{7Li}$ uncertainty, this is mainly due to the $^4He +
{^3He}\lrt \gamma + {^7Be}$ rate, with $\omb$ contributing for
$\sim 50\%$ to the total error on $X_{7Li}$. We show in Table
\ref{table:contrsigmaheli} the relative contribution of the most
important rates to $(\sigma^{th}_{44})^2$ and
$(\sigma^{th}_{77})^2$.

\begin{table}[b]
\begin{center}
\bt{|c|c|c|c|} \hline\hline rate  $R_k$ & $(\delta
\sigma^{th}_{44k}/\sigma^{th}_{44})^2$(\%) & rate $R_k$ & $(\delta
\sigma^{th}_{77k}/\sigma^{th}_{77})^2$(\%) \\
\hline\hline $n \lrt p$ & 29 & $^4He+{^3He} \lrt \gamma+{^7Be}$ &
68 \\ $D+D\lrt n+{^3He}$ & 29 & $^3He+D\lrt p+{^4He}$ & 9
\\ $D+D \lrt p+{^3H}$ & 26 & $D+D \lrt n+{^3He}$ & 8.5\\ $p+n\lrt \gamma+D$ & 4
& $^7Be+n\lrt {^4He}+^4He$ & 7 \\
\hline\hline
\et
\end{center}
\bigskip
\caption{The main contributions to $(\sigma^{th}_{44})^2$ and
$(\sigma^{th}_{77})^2$, in percentage, for $\neff=3.01$ and
$\omb=0.023$. $n \lrt p$ denotes the weak processes.}
\label{table:contrsigmaheli}
\end{table}

The status of the primordial abundance measurements for ${^4He}$
and ${^7Li}$ is still quite involved. There are different
determinations of $Y_p$, $Y_p=0.234 \pm 0.003$ \cite{lowyp} (low
$Y_p$), $Y_p=0.244 \pm 0.002$ \cite{it} (high $Y_p$), and
$Y_p=0.2421 \pm 0.0021$ \cite{it03}. The latter is the result of a
recent analysis where the authors performed an accurate study of
systematics. If the statistical error were not underestimated,
these values would be only marginally compatible. Waiting for new
experimental data, we adopt in our analysis a conservative point
of view by using
\be
Y_p = 0.239 \pm 0.008\, .
\label{conservative}
\ee

\begin{figure}[t]
\begin{center}
\epsfig{file=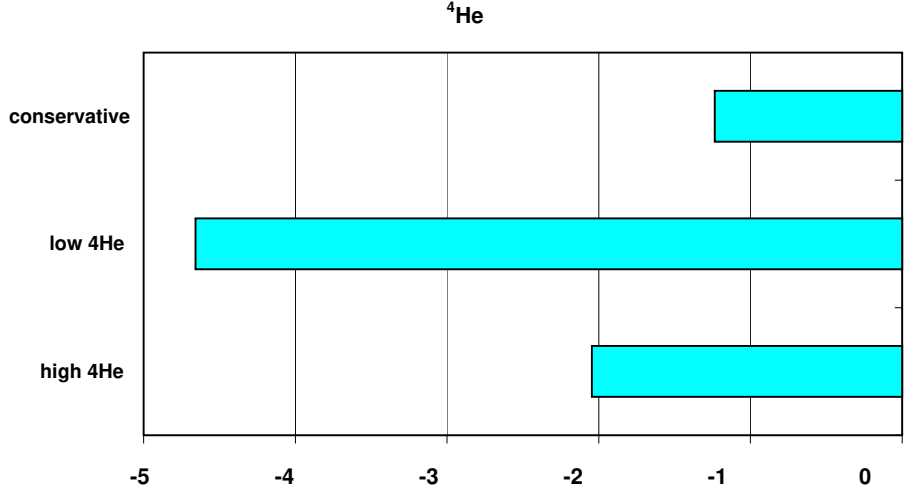,width=12truecm}
\end{center}
\caption{The pulls (see text) of $Y_p$ measurements with respect
to the theoretical prediction for $\neff=3.01$ and $\omb=0.023$,
in units of $\left( (\sigma^{th}_{44})^2 + (\sigma^{exp}_4)^2
\right)^{1/2}$.}
\label{fig:pullhe}
\end{figure}

As it is evident from Fig.~\ref{fig:pullhe}, where we report the
pulls with respects to expected theoretical $Y_p$ in unit of the
combined error $\left( (\sigma^{th}_{44})^2 + (\sigma^{exp}_4)^2
\right)^{1/2}$, even the higher value of $Y_p$ \cite{it} appears
in moderate disagreement ($2 \sigma$ effect) with standard BBN.
If we let $\neff$ to be a free parameter it is therefore not a
surprise that $Y_p$ favors a slightly lower value for $\neff$. In
fact, a way to decrease $Y_p$ is to delay the freezing out of weak
rates converting neutron and protons, and this can be achieved by
reducing the number of relativistic species contributing to the
Hubble parameter. In our opinion much effort should be devoted to
have a more clear understanding of the experimental results or to
undertake a new measurement campaign for ${^4He}$. On the
theoretical side, in fact, we may say that the value of $Y_p$ is
quite robust, and affected by a very small total error. New data
would eventually tell us whether the standard BBN is fully
consistent, or more exotic scenarios should be invoked. We discuss
one of these, a degenerate neutrino distribution, in next Section.

\begin{figure}[t]
\begin{center}
\epsfig{file=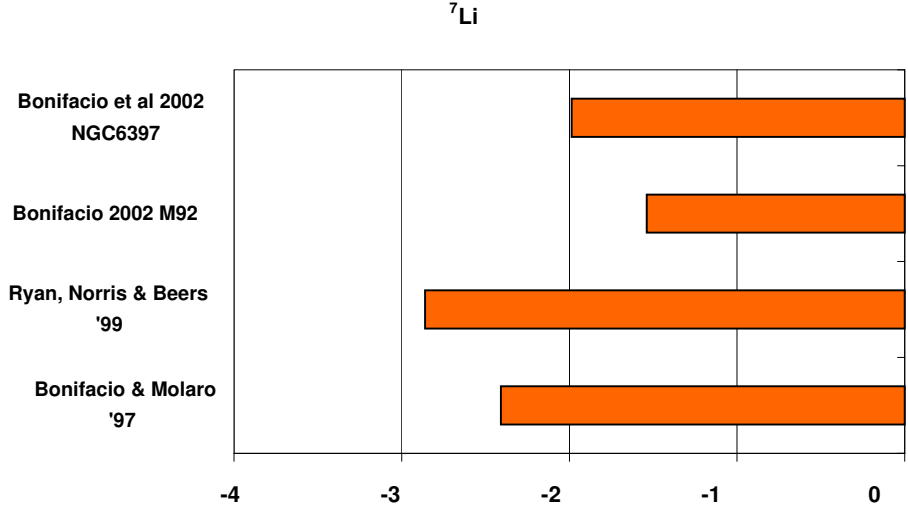,width=12truecm}
\end{center}
\caption{The pulls (see text) of $X_{7Li}$ measurements with
respect to the theoretical prediction for $\neff=3.01$ and
$\omb=0.023$, in units of $\left((\sigma^{th}_{77})^2 +
(\sigma^{exp}_7)^2 \right)^{1/2}$.}
\label{fig:pullli}
\end{figure}

Concerning $^7Li$ experimental measurements
\cite{bm}-\cite{b20032}, the primordial origin of the Spite
plateau has been recently questioned. In particular the authors of
\cite{ryan} found evidence for a dependence of $X_{7Li}$ on
metallicity. In any case, there are some indication for a
depletion mechanism of $^7Li$ observed in POP II metal poor stars.
However, as can be seen in Fig.~\ref{fig:pullli}, the discrepancy
between the most recent observations of
\cite{b20031}-\cite{b20032} and our theoretical value is now
reduced to a less than $3\sigma$ effect. The average over the
different observed values for $X_{7Li}$ \cite{bm}-\cite{b20032},
which are mutually compatible, gives $X_{7Li} = (1.70 \pm
0.17)\times 10^{-10}$.

\begin{figure}[t]
\begin{center}
\epsfig{file=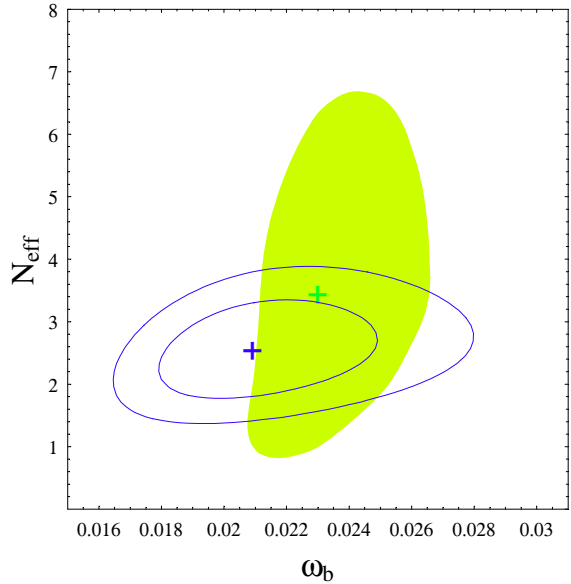,width=8truecm}
\end{center}
\caption{The 68 and $95\%$ C.L. contours for the $D+^4He$
likelihood function in the $\omb-\neff$ plane. We also show the
result of the $CMB+D$ analysis of Fig.~\ref{fig:likelihoodcmbd}
(left plot).}
\label{fig:likelihoodbbn}
\end{figure}
As in previous studies we do not use $X_{7Li}$ in our BBN
likelihood analysis, waiting for a more clear understanding of a
possible depletion mechanism. Moreover, further efforts should be
also devoted in trying to reduce the large theoretical uncertainty
which is still affecting $X_{7Li}$, mainly by lowering the
uncertainty on the leading reaction $^4He + {^3He} \lrt \gamma +
{^7Be}$, presently of the order of $18\%$. As we mentioned, this
produces a large fraction of the overall $20\%$ error on
$X_{7Li}$.

Using the $X_D$ result of \cite{deuterium} and Eq.
(\ref{conservative}) for $Y_p$, we consider the likelihood
function as defined in (\ref{likelibbn}), as function of $\neff$
and $\omb$. Our results for 68 and $95\%$ C.L. contours are shown
in Fig.~\ref{fig:likelihoodbbn}, where for comparison we also
report the $95\%$ C.L. contour of the $CMB+D$ analysis. The effect
of $^4He$ has been to shift both $\neff$ and $\omb$ towards
smaller values. In particular, after marginalization, we get at
$2\sigma$, $\omb=0.021^{+0.005}_{-0.004}$ and
$\neff=2.5^{+1.1}_{-0.9}$. Notice how, with present determination
of $Y_p$ and $X_D$, the range obtained for $\omega_b$ is rather
broad compared with WMAP result.

We close this section with the theoretical estimate on $^3He$ and
$^6Li$ number fractions. For $\omb=0.021$ and $\neff=2.5$ we get
$X_{3He}= (1.02^{+0.10}_{-0.11})\cdot 10^{-5}$, which saturates
the recent estimate reported in \cite{bania} for the upper limit
of the $^3He$ to $H$ number density from observations of $HII$
regions and planetary nebulae, $X_{3He}=(1.1 \pm 0.2)\cdot
10^{-5}$. The $^6Li$ primordial density is predicted to be quite
small, $X_{6Li} = (1.2 \pm 1.7)\cdot 10^{-14}$, which results in a
ratio $^6Li/{^7Li}=(2.8\pm 4.0)\cdot 10^{-5}$, much smaller than
the experimental value $^6Li/{^7Li}=0.05$, measured in two stars
in the galactic halo (see, for example, \cite{6li} and refs.
therein). Very close result are obtained for the other analysis
considered in the previous Section. We recall however that the
theoretical estimate is still suffering the large uncertainty
affecting the rate $^4He+D\lrt \gamma+{^6Li}$, which is supposed
to give the main production channel during BBN (see our discussion
in Section 2). Actually any $^6Li$ detection has important
consequences in constraining the destruction rate of $^7Li$ in POP
II stars, as well as exotic processes to $D$ synthesis, as for
example photodissociation or spallation of $^4He$ \cite{jedamzik}.
A summary of the ranges for $\omb$ and $\neff$ and of nuclei
yields is reported in Table 4.

\begin{table}[p]
\begin{center}
\bt{|c|c|c|c|} \hline\hline  & $ \begin{array}{c} CMB+BBN(D)
\\at\, N_{eff}=3.01
\end{array} $
 & $CMB+BBN(D)$
& $BBN(D+{^4He})$\\
\hline\hline $
\begin{array}{c}
\\ \omega_b\,(2\,\sigma)
\\{}
\end{array}$ & $0.023^{+0.003}_{-0.002}$ & $0.023^{+0.003}_{-0.002}$ &
$0.021^{+0.005}_{-0.004}$ \\
\hline $
\begin{array}{c}
\\N_{eff}\,(2\,\sigma)
\\{}
\end{array}$ & 3.01 & $3.6^{+2.5}_{-2.3}$ & $2.5^{+1.1}_{-0.9}$ \\
\hline $
\begin{array}{c}
\\X_D\cdot 10^5
(1\,\sigma)\\ (2.78^{+0.44}_{-0.38})\\{}
\end{array}$ & $2.55^{+0.22}_{-0.34}$ & $2.74^{+0.46}_{-0.51}$ &
$2.75^{+0.53}_{-0.57}$ \\
\hline $
\begin{array}{c}
 \\
 Y_p\, (1\,\sigma)\\ (0.239\pm 0.008)\\{}
\end{array}$ & $0.2483^{+0.0008}_{-0.0005}$ &
$0.256^{+0.014}_{-0.015}$ & $0.240 \pm 0.008$\\
\hline $
\begin{array}{c}
\\X_{7Li}\cdot 10^{10}\, (1\,\sigma)\\ (1.70 \pm 0.17)\\{}
\end{array}$ &$4.9^{+1.4}_{-1.2}$ & $4.7^{+1.5}_{-1.3}$ &
$4.3^{+1.6}_{-1.3}$\\
\hline $
\begin{array}{c}
\\X_{3He}\cdot 10^{5}\, (1\,\sigma)\\ (1.1\pm 0.2)\\{}
\end{array}$ & $0.99^{+0.07}_{-0.08}$ & $1.01^{+0.09}_{-0.10}$ &
$1.02^{+0.10}_{-0.11}$\\
\hline $
\begin{array}{c}
\\X_{6Li}/X_{7Li}\,(1\,\sigma)\\ (0.05)\\{}
\end{array}$ & $(2.4 \pm 3.4)\cdot 10^{-5}$ & $(2.8\pm 3.7)\cdot
10^{-5}$ & $(2.8\pm 4.0)\cdot 10^{-5}$\\
\hline \et
\end{center}
\bigskip
\caption{A summary of the results obtained in the three different
SBBN likelihood analysis discussed in this work. For nuclear
abundances we report in parenthesis the experimental value or the
best estimate currently available: actually, only in some cases a
direct comparison is possible, as explained in the text. The
larger uncertainties affecting $Y_p$ in the third and fourth
columns are due to the variation of $\neff$ in the $1-\sigma$
range.} \label{summarytable}
\end{table}

\section{Bounds on neutrino chemical potentials.
Still room for extra relativistic species} \setcounter{equation}0
\noindent

The cosmological effects of non zero neutrino antineutrino
asymmetries have received a renewed attention in the last few
years \cite{hansenetal}, \cite{riotto}-\cite{orito}. The bounds
from BBN and CMB data on the asymmetry parameters $\xi_{\alpha}$
for the three active neutrinos, $\alpha=e,\mu,\tau$, are more
stringent by considering the effect of oscillations in primordial
universe. In fact, in \cite{pastor2} it was shown that for the
large mixing angle solution to solar neutrino problem, presently
favored by data, oscillations tend to homogenize the $\xi_\alpha$,
which are therefore all bounded by the more stringent constraint
on $\xi_e$. Neutrino chemical potentials affect BBN via their
contribution to energy density, \be \rho_\nu = \frac{7}{8}\,
\left( \frac{T_\nu}{T_\gamma} \right)^{4}~ \rho_\gamma~ \left\{
3.01 + \sum_\alpha \left[ \frac{30}{7}
\left(\frac{\xi_\alpha}{\pi}\right)^2 + \frac{15}{7}
\left(\frac{\xi_\alpha}{\pi}\right)^4 \right] \right \}\, ,
\label{rhonudeg} \ee and so to Hubble parameter. In particular
$\xi_{\mu,\tau}$ only enters in changing the value of $\neff$. On
the other hand $\xi_e$ also affects the $n/p$ chemical equilibrium
at freeze out, changing the neutron to proton ratio as
$\exp(-\xi_e)$. A positive $\xi_e$ (more neutrinos than
antineutrinos) favors $n\rt p$ reactions with respect to inverse
processes, and this eventually translates into a smaller value for
$Y_p$.

In this section, using the WMAP result on $\omb$ as a prior, we
use our BBN code to update our previous analysis \cite{emmpjhep}
of degenerate BBN (DBBN), with two main purposes:
\begin{itemize}
\item[i)] since, as we stressed in the previous Section, the $Y_p$
experimental value is slightly lower than what expected in
standard BBN, we analyze how the agreement improves allowing for
non zero $\xi_\alpha$, but fixing the number of relativistic
species, besides photons, to the three active light neutrinos.
\item[ii)] in the general case we determine the upper bounds in DBBN on
$\neff$ and $\xi_\alpha$.
\end{itemize}

We construct the BBN likelihood function ${\cal L}_{BBN}
(\neff,\omb,\xi_e)$ defined in Eq.~(\ref{likelibbn}), with the
extra dependence on the parameter $\xi_e$. The use of $Y_p$ may
seem inconsistent with our discussion of the previous Section. For
standard BBN in fact we stressed that the experimental value of
$Y_p$ shows a slight disagreement with the theoretical estimate,
possibly a sign of unaccounted systematic effects. However the
introduction of a (positive) $\xi_e$ in the DBBN case leads to a
smaller theoretical value for $Y_p$ even for $\neff\sim 3$, so in
this case one may interpret the experimental findings as an
indication of a neutrino-antineutrino asymmetry.

\begin{figure}[t]
\bt{cc}
\epsfig{file=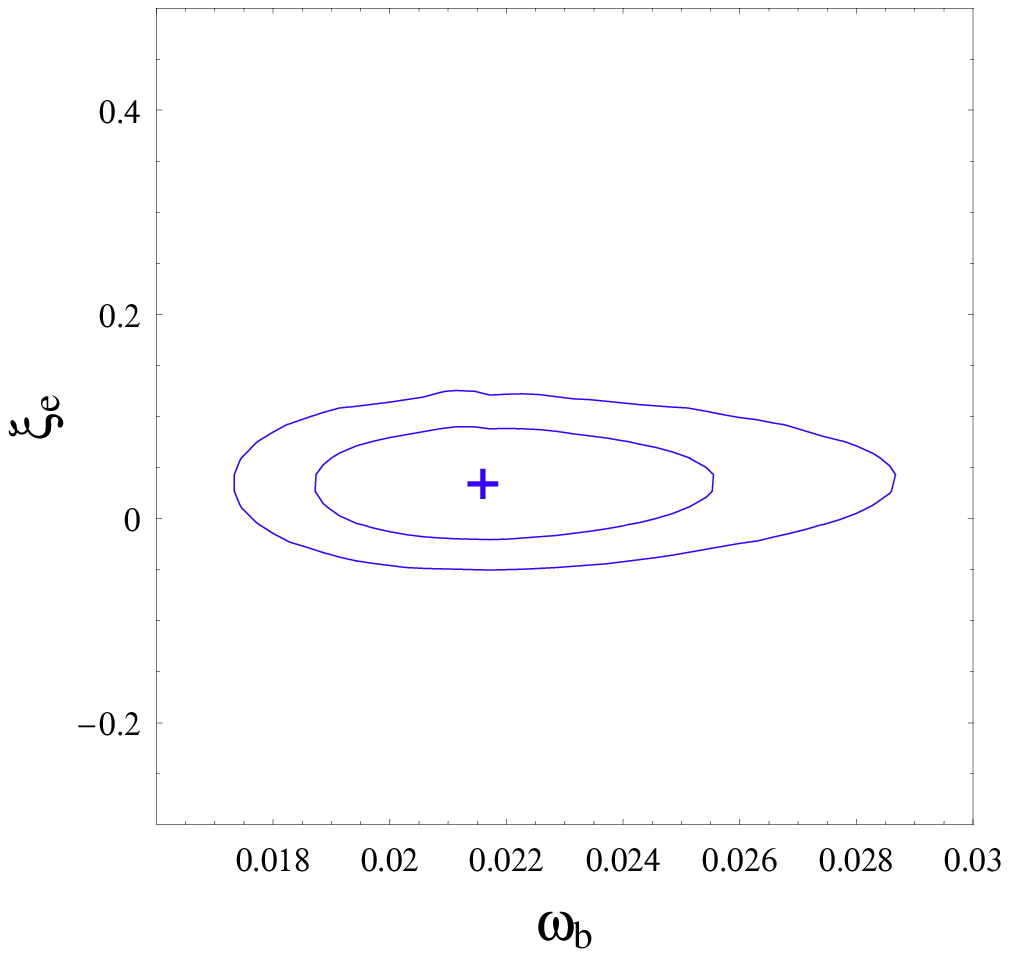,width=6.5truecm} &
\epsfig{file=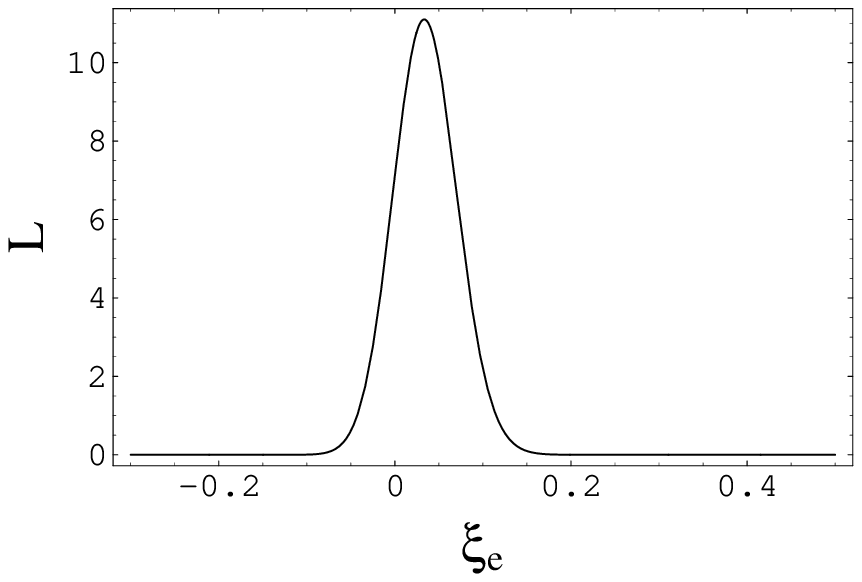,width=8truecm}
\et
\caption{The 68 and $95\%$ contours for the $D+^4He$ degenerate
likelihood function in the $\omb-\xi_e$ plane in the scenario with
only three active neutrinos and without prior on $\omb$ (left
plot). Right plot shows the marginalized likelihood versus $\xi_e$
when a prior on $\omb$ is imposed, $\omega_b=0.023\pm 0.002$.}
\label{fig:onlyxie}
\end{figure}
In the minimal scenario we assume only three active neutrinos
contributing to $\neff$. Since $\xi_{\mu,\tau}$ are enforced by
oscillations to be of the same order of magnitude of $\xi_e$, we
can write down a constraint for $\neff$, \be \neff = 3.01 + 3
\left[ \frac{30}{7} \left(\frac{\xi_e}{\pi}\right)^2 +
\frac{15}{7} \left(\frac{\xi_e}{\pi}\right)^4 \right] = \bar \neff
(\xi_e)\, . \label{neff1} \ee In the second scenario we
parameterize possible extra light degrees of freedom with
$\Delta\neff$, defined by \be \neff = \bar \neff (\xi_e) +
\Delta\neff\, . \label{neff2} \ee In both these cases we adopt the
experimental values/errors for $Y_p$ and $X_D$ reported in Section
4 and 5.

\begin{figure}[t]
\begin{center}
\epsfig{file=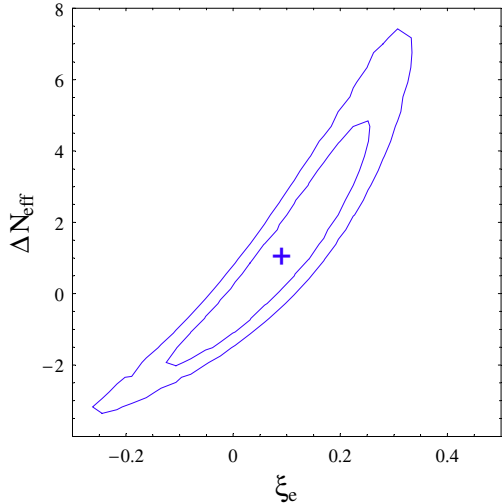,width=7truecm}
\end{center} \caption{The 68 and $95\%$ C.L. contours in the
$\xi_e-\Delta\neff$ plane for the degenerate BBN likelihood
function.}
\label{fig:likelihoode}
\end{figure}

We show in Fig.~\ref{fig:onlyxie} (left plot) the $68$ and $95\%$
C.L. contours for the first scenario. When imposing the WMAP prior
with a conservative error ($\omega_b=0.023\pm 0.002$) we obtain
the marginalized likelihood for $\xi_e$ reported in the right plot
of Fig.~\ref{fig:onlyxie}, giving $\xi_e = 0.03^{+0.08}_{-0.06}$
(2$\sigma$). Correspondingly $X_D=2.51\cdot 10^{-5}$,
$Y_p=0.2414$.

Fig.~\ref{fig:likelihoode} summarizes our findings in the second
scenario. At $95\%$ C.L., we get $\neff=3.9^{+4.8}_{-3.4}$ and
$\xi_e = 0.13^{+0.18}_{-0.26}$, again using the WMAP prior on
$\omega_b$. We notice that our limits on $\neff$ are broader than
what found in \cite{barger2}, since we use only the $\omega_b$
prior from CMB in the analysis. This results in a more
conservative bound, independent on the assumption that $\neff$ is
the same at both the CMB and BBN epochs. Our upper limit on
$\neff$, being sensibly larger than 3, should be understood as a
direct bound to possible extra light particles. Notice that in the
DBBN scenario a fourth sterile neutrino, as for example required
to interpret LSND evidence for $\ov{\nu}_\mu \lrt \ov{\nu}_e$
oscillation, is not yet ruled out. See also \cite{hannestad} for a
neutrino mass analysis of this issue.

\section{Conclusions}
\setcounter{equation}0 \noindent

In this paper we have discussed the present status of Primordial
Nucleosynthesis after the recent measurements of CMB anisotropies
by the WMAP experiment, which represents a new relevant step
towards a better understanding of the evolution of the universe,
and in particular gives a very precise determination of the baryon
density parameter $\omb$. The question of whether the standard
picture of BBN is a satisfactory scenario for production of light
nuclei requires an increased precision in both its theoretical and
experimental aspects.

As discussed in the paper, it seems to us that some results in
this direction have already been achieved. The accuracy of
theoretical estimates has been improved by refining the estimate
of $n/p$ ratio at decoupling. Most importantly, a complete
reanalysis of all nuclear rates which enter the BBN network has
been made, using the NACRE compilation, as well as all most recent
available results. This analysis has been used to implement a new
BBN numerical code. Particular attention has been devoted to the
rates which mainly contribute to evolution of observed nuclei
abundances. Moreover several processes which were not considered
in previous analysis have been added. The main outcome of this
study is a reduced theoretical uncertainty of the BBN model, which
is presently well below the corresponding experimental errors, at
least for $D$ and $^4He$. The $^7Li$ number fraction is still
affected by a $20\%$ uncertainty, mainly due to the effect of
$^4He +{^3He} \lrt \gamma +{^7Be}$ rate. More accurate
measurements of this rate are therefore strongly desirable.

From the experimental point of view, Deuterium determinations show a
very good agreement with the corresponding theoretical expectation,
and are fully compatible with the value of $\omb$ suggested by WMAP
data. This is quite remarkable, since $D$ is strongly dependent on the
baryon content of the universe. On the other hand we do think that new
measurements of $^4He$ and $^7Li$ should be performed, along with a
clear understanding of possible depletion mechanisms of primordial
$X_{7Li}$. It should be mentioned however that the most recent results
for $^7Li$, compared with our theoretical prediction, reduce the
disagreement for $X_{7Li}$ at less than $3\sigma$ level. The $^4He$
mass fraction still has two distinct determinations which are mutually
incompatible. Even adopting a conservative value, as we did in this
paper, still the amount of $^4He$ experimentally detected seems
slightly lower than expected for a standard BBN scenario. This may be
seen either as due to possible systematics in the measurements and/or
extrapolation technique for $Y_p$, or rather as the effect of exotic
physics in the BBN scenario, as for example neutrino degeneracy. For
an alternative method, still affected by large uncertainty, to infer
$Y_p$ see Ref. \cite{newhansen}. Possible measurements of other
nuclei, like $^3He$, for which there is a single recent estimate, or
$^6Li$, though difficult\footnote{Actually, only for a smaller
$\omega_b$, a large $^4He+D\rt \gamma+ {^6Li}$ reaction rate, at the
upper limit of the present uncertainty range, and almost no depletion
mechanism in act in hot and metal-poor PopII halo stars, one could
reveal the
\emph{primordial} (as opposed to the cosmic-ray produced) $^6{Li}$ in
the near future.}, may greatly help in clarifying the overall
soundness of Primordial Nucleosynthesis scenarios.

The main results of this paper are reported in Table 4 while our
analysis can be summarized as follows:
\begin{itemize}
\item[i)] we have described some key examples of the study
performed to improve the accuracy of the nuclear network adopted
in our code. A full description will be presented elsewhere;
\item[ii)] we have performed a likelihood analysis of the WMAP
results and BBN (Deuterium) abundance on $\omb$ and $\neff$,
showing that the simplest standard BBN, with three neutrino as the
only relativistic particles besides photons at the BBN epoch is
largely consistent. In particular we found, at $95\%$ C.L.
$\omb=0.023^{+0.003}_{-0.002}$ and $\neff=3.6^{+2.5}_{-2.3}$.
Notice the weak sensitivity of both $D$ and CMB data to $\neff$.
We stressed however that an improved accuracy of $D$ measurement
may provide more accurate bounds on this parameter, {\it
independently} of information on $Y_p$; \item[iii)] we have
studied, again via a likelihood method of the $X_D$ and $Y_p$
observables, the theory versus experiment status. We find that
taking into account the present determination of $^4He$ shifts the
preferred values of both $\neff$ and $\omb$ towards smaller values
than in the $CMB+D$ analysis, $\omb=0.021^{+0.005}_{-0.004}$ and
$\neff=2.5^{+1.1}_{-0.9}$ at $95\%$ C.L.. Finally we have also
reported our estimate for $^3He$ fractional density and
$^6Li/{^7Li}$ ratio (see Table 4); \item[iv)] in the framework of
degenerate BBN we reported the bounds on possible extra
relativistic degrees of freedom. Oscillations enforce the three
active neutrinos to share the same small chemical potential at the
BBN epoch and we find as a common bound $-0.03\leq
\xi_{e,\mu,\tau}\leq 0.11$ at $95\%$ C.L. for $\neff$
corresponding to three active neutrinos. In this case the
agreement of $^4He$ theoretical expectation with experimental data
improves. If we leave $\neff$ as a free parameter we get $\neff
=3.9^{+4.8}_{-3.4}$ at $95\%$ C.L.. In particular the upper bound
represents the largest extra contribution to the relativistic
energy density due to exotic particles consistent with BBN.
\end{itemize}

After the huge progresses in the study and detection of Cosmic
Microwave Background anisotropies, we think that further efforts
in refining the BBN theoretical and experimental aspects may
provide new pieces of information on cosmology and fundamental
physics.

\noindent {\large \bf Acknowledgements}\\ The authors are grateful
to V.B. Belyaev, G. Imbriani, A. Melchiorri, S. Pastor, C. Rolfs,
and F. Terrasi for useful discussions and valuable comments.

\end{document}